\pdfoutput=1
\documentclass[iop]{emulateapj}
\usepackage{lineno}
\def\farcs{\hbox{$.\!\!^{\prime\prime}$}}

\slugcomment{}
\shorttitle{The disk-outflow system around the rare young O-type protostar W42-MME}
\shortauthors{L.~K. Dewangan et al.}
\begin{document}
\title{The disk-outflow system around the rare young O-type protostar W42-MME}
\author{L.~K. Dewangan\altaffilmark{1}, I. I. Zinchenko\altaffilmark{2}, P. M. Zemlyanukha\altaffilmark{2}, 
S.-Y. Liu\altaffilmark{3}, Y.-N. Su\altaffilmark{3}, S. E. Kurtz\altaffilmark{4}, D.~K. Ojha\altaffilmark{5},\\ 
A. G. Pazukhin\altaffilmark{2}, and Y.~D. Mayya\altaffilmark{6}}
\email{lokeshd@prl.res.in}
\altaffiltext{1}{Physical Research Laboratory, Navrangpura, Ahmedabad - 380 009, India.}
\altaffiltext{2}{Institute of Applied Physics of the Russian Academy of Sciences 46 Ul'yanov~str., 603950 Nizhny Novgorod, Russia.}
\altaffiltext{3}{Institute of Astronomy and Astrophysics, Academia Sinica P.O. Box 23-141, Taipei 10617, Taiwan, R.O.C.}
\altaffiltext{4}{Instituto de Radioastronomía y Astrof\'{i}sica, Universidad Nacional Aut\'{o}noma de M\'{e}xico, Apartado Postal.}
\altaffiltext{5}{Department of Astronomy and Astrophysics, Tata Institute of Fundamental Research, Homi Bhabha Road, Mumbai - 400005, India.}
\altaffiltext{6}{Instituto Nacional de Astrof\'{i}sica, \'{O}ptica y Electr\'{o}nica, Luis Enrique Erro \# 1, Tonantzintla, Puebla, M\'{e}xico C.P. 72840.}
\begin{abstract}
We present line and continuum observations (resolution $\sim$0\rlap.{$''$}3--3\rlap.{$''$}5) made with 
the Atacama Large Millimeter/submillimeter Array (ALMA), Submillimeter Array, and Very Large Array of 
a young O-type protostar W42-MME (mass: 19$\pm$4 M$_{\odot}$). 
The ALMA 1.35 mm continuum map (resolution $\sim$1$''$) shows that 
W42-MME is embedded in one of the cores (i.e., MM1) located within a thermally supercritical 
filament-like feature (extent $\sim$0.15 pc) containing three cores (mass $\sim$1--4.4 M$_{\odot}$). 
Several dense/hot gas tracers are detected toward MM1, suggesting the presence 
of a hot molecular core with the gas temperature of $\sim$38--220~K. 
The ALMA 865 $\mu$m continuum map (resolution $\sim$0\rlap.{$''$}3) reveals at least five continuum sources/peaks (``A--E") within 
a dusty envelope (extent $\sim$9000~AU) toward MM1, where shocks are traced in the SiO(8--7) emission.
The source ``A" associated with W42-MME is seen almost at the center of the dusty envelope, and is surrounded by other continuum peaks. 
The ALMA CO(3--2) and SiO(8--7) line observations show the bipolar outflow extended below 10000 AU, which is driven by the source ``A". 
The ALMA data hint the episodic ejections from W42-MME. 
A disk-like feature (extent $\sim$2000~AU; mass $\sim$1 M$_{\odot}$) with velocity gradients is investigated in the source ``A" (dynamical mass $\sim$9 M$_{\odot}$) using the ALMA H$^{13}$CO$^{+}$ emission, and is perpendicular to the CO outflow. 
A small-scale feature (below 3000 AU) probably heated by UV radiation from the O-type star is also investigated toward the source ``A".
Overall, W42-MME appears to gain mass from its disk and the dusty envelope. 
\end{abstract}
\section{Introduction}
\label{sec:intro}
Unraveling the exact formation mechanism of massive OB-type stars (M $\gtrsim$ 8 M$_{\odot}$) is one of the outstanding 
issues in massive star formation (MSF) research. It is directly related to the understanding of the process of mass accumulation in MSF, which is also a key open research problem. Both theoretical and observational studies of the birth process of massive stars have been extensively performed, and face serious difficulties. These aspects are thoroughly discussed in numerous review 
articles \citep[e.g.,][]{mckee07,zinnecker07,krumholz09,krumholz12,tan14,Motte+2018,hirota18,rosen20}. 
Massive stars are often located in embedded and crowded environments, associated with outflows and jets, and seen at junctions of dust and molecular filaments (i.e., hub-filament systems).
Based on these observational features, five major theoretical scenarios have been discussed in the literature to 
explain the formation of massive stars; (1) the turbulent core (TC)/core accretion/ monolithic collapse model \citep{mckee03}; (2) the competitive accretion (CA) model \citep{bonnell02,bonnell04,bonnell06}; 3) the global hierarchical collapse (GHC) model \citep{Vazquez-Semadeni+2009,Vazquez-Semadeni+2017,Vazquez-Semadeni+2019}; 
4) the global non-isotropic collapse (GNIC) scenario \citep{Tige+2017,Motte+2018}; 5) the inertial-inflow model \citep{padon20}.

In the TC model, a massive star or a small number of multiples can form through the 
collapse of a massive, isolated, and gravitationally bound prestellar core, which is supported by magnetic and/or supersonic turbulent pressures. 
The monolithic collapse is considered as an extension of the model of formation of low-mass stars 
to the massive ones, but with higher accretion rates. 
According to the CA model, a mass assembly is achieved via global gravitational forces in the central part of 
the clumps confined by smaller scale multiple cores. 
In other words, a massive star can form via rapid growth of low-mass protostellar seeds by mass accretion from surrounding gas. In the CA model, the position of more massive stars is predicted at the centre of a protostellar cluster. 
In the GHC scenario, gravitationally driven fragmentation operates in star-forming molecular clouds, and large scale accretion flows are expected to directly fed 
massive star-forming regions \citep[see also][]{rosen20}. In the GNIC model, \citet{Motte+2018} presented an empirical scenario for the formation of massive stars, which uses the flavours of the GHC and clump-feed accretion scenarios \citep[see][]{Vazquez-Semadeni+2009,Vazquez-Semadeni+2017,Smith+2009}. In this model, the massive protostellar cores can form from low-mass protostellar cores, which accrete further material from their parental massive dense core (MDC). 
Concerning such investigation, one requires the identification of a hub-filament system containing the highest density regions, 
where multiple accreting filaments converge.
According to the inertial-inflow model, a massive star can form via large scale, converging, inertial flows, which are originated due to supersonic turbulence \citep{padon20}. In this model, the gravity of the star does not control the inertial inflow, which is driven by large-scale turbulence. 
These authors also pointed out that a massive star cannot form through the collapse of massive cores (i.e., TC model) and the CA process. In the inflow region, one expects the smaller inflow velocity than the turbulent velocity. Hence, one can observationally compare the inflow and turbulent velocity components to check the applicability of this model \citep[e.g.,][]{padon20}.

In order to observationally assess the aforementioned theoretical scenarios, one needs to study the complex circumstellar structures of genuine massive young 
stellar objects (MYSOs) and the physical properties of their parental cores using a multi-scale and multi-wavelength approach. 
It also requires the knowledge of the kinematics of the dense gas toward MYSOs including 
their inner circumstellar structures because such objects are believed to hold the initial condition of MSF. 
In this relation, the present paper focuses on a genuine MYSO located within the larger massive star-forming complex W42 \citep{dewangan15b}. 
The target MYSO is associated with the 6.7 GHz methanol maser emission (MME).

W42 hosts a bipolar H\,{\sc ii} region \citep[e.g.,][]{woodward85,lester85,anderson09,dewangan15c} and a 6.7 GHz MME \citep[radial velocity (V$_\mathrm{lsr}$) $\sim$58.1 km s$^{-1}$;][]{szymczak12}. Based on the near-infrared (NIR) photometric and spectroscopic observations, \citet{blum00} found that the W42 H\,{\sc ii} region is powered by an O5-O6 star. Observations of the C\,{\sc ii} and $^{3}$He radio recombination lines show a radial velocity of the ionized gas in the W42 H\,{\sc ii} region to be $\sim$59.6 km s$^{-1}$ \citep{quireza06}. 
The molecular cloud associated with W42 \citep[i.e., U25.38$-$0.18;][]{anderson09} has been studied in a velocity 
range of [58, 69] km s$^{-1}$ \citep[see also][]{dewangan15c}. 
A very similar observed velocity of the ionized gas and the molecular gas suggests that the W42 H\,{\sc ii} region and the 6.7 GHz MME belong to the same physical system. 
A distance of 3.8 kpc to W42 has been adopted in the literature \citep[e.g.,][]{lester85,anderson09,dewangan15b,dewangan15c}. 

\citet{dewangan15b} found the position of the 6.7 GHz MME at the center of a 
parsec-scale bipolar outflow in the H$_\mathrm{2}$ image. 
They also investigated an infrared counterpart (IRc) of 6.7 GHz MME in W42 (i.e., W42-MME) using the infrared images at wavelengths longer than 2.2~$\mu$m (see also Figure~1 in their paper). 
W42-MME has been characterized as a rare O-type protostar (mass: 19$\pm$4 M$_{\odot}$ and visual extinction: 48$\pm$15 mag) 
with a luminosity of $\sim$4.5 $\times$ 10$^{4}$ L$_{\odot}$ \citep{dewangan15b}. At the sensitivity of the Coordinated Radio and Infrared Survey for High-Mass Star Formation \citep[CORNISH;][]{hoare12} 5~GHz continuum map (resolution $\sim$1\rlap.{$''$}5; rms $\sim$0.4 mJy beam$^{-1}$), no radio counterpart of the MYSO W42-MME was detected \citep[e.g.,][]{dewangan15b}.   
Based on these findings, \citet{dewangan15b} proposed this object as a genuine MYSO in a very early evolutionary stage, prior to an ultracompact (UC) H\,{\sc ii} phase. 

\citet{dewangan15b} also examined the inner environment of this object using European Southern Observatory (ESO) Very Large Telescope (VLT) NAOS-CONICA (NACO) near infrared (NIR) adaptive-optics images at K$_{s}$-band ($\lambda$ = 2.18 $\mu$m; resolution $\sim$0\rlap.{$''$}2 or 760~AU at a distance of 3.8~kpc) and L$^{\prime}$-band ($\lambda$ = 3.8 $\mu$m; resolution $\sim$0\rlap.{$''$}1 or 380~AU). The VLT/NACO L$^{\prime}$ image 
allowed them to investigate an infrared envelope/outflow cavity (extent $\sim$10640~AU) containing a point-like source and a collimated jet-like feature. This point-like source has been proposed as the main powering source of the infrared jet and outflow. 
They also found that the infrared envelope/outflow cavity was tapered at both ends, and was aligned along the north-south direction. 
Along the flow axis, two blobs with diffuse emission have been traced in the NACO image, and were located at a similar distance 
of $\sim$11800~AU from the main powering source \citep[see][for more details]{dewangan15b}. 
Based on the NIR polarimetric study carried out by \citet{jones04}, 
\citet{dewangan15b} suggested that the outflow axis traced in the H$_{2}$ map is parallel to the magnetic field at the position 
angle of $\sim$15$\degr$. 

Due to coarse beam sizes (i.e., 16$''$--46$''$), the publicly available surveys of molecular line data (i.e., the FOREST Unbiased Galactic plane Imaging survey with the Nobeyama 45-m telescope \citep[FUGIN;][]{umemoto17}, the CO High-Resolution Survey \citep[COHRS;][]{dempsey13}, and 
the Galactic Ring Survey \citep[GRS;][]{jackson06}) cannot resolve the jet-outflow system as traced in the infrared images. Hence, the molecular content of the promising infrared jet-outflow system in W42-MME (hereafter jet-outflow W42-MME system) is not yet known. Furthermore, we do not know the physical properties or the spatial morphology of the clump/core containing W42-MME. 

In this paper, we explore the inner circumstellar environment (1000--10000 AU scales) of the O-type protostar (W42-MME) using the multi-scale and multi-wavelength continuum and line data sets (resolutions $\sim$0\rlap.{$''$}3--3\rlap.{$''$}5), which were obtained from the Submillimeter Array (SMA), Very Large Array (VLA), and Atacama Large Millimeter/submillimeter Array (ALMA). These data sets have been analyzed to examine in detail the morphological and kinematical structure of the molecular gas immediately
associated with and surrounding the rare jet-outflow system near the O-type protostar. It is possible because the resolution of the ALMA continuum and line data ($\sim$0\rlap.{$''$}3) is almost similar to that of the previously published VLT/NACO L$^{\prime}$ image ($\sim$0\rlap.{$''$}1). The ALMA data also allow us to explore the physical properties, spatial morphology, and kinematics of the core hosting W42-MME. The outcomes derived using high spatial resolution continuum maps and different density tracer lines provide an opportunity to assess the existing wide range of theoretical MSF models as highlighted earlier in this section. 

In Section~\ref{sec:obser}, we summarize the observations. 
In Section~\ref{sec:data}, we present the results concerning the physical environment of W42-MME. In this section, we also 
discuss core properties and core kinematics.
In Section~\ref{sec:disc}, we discuss possible star formation processes in W42-MME. 
Finally, in Section~\ref{sec:conc}, we summarize our main conclusions.
\section{Data sets and analysis}
\label{sec:obser}
\subsection{Atacama Large Millimeter/submillimeter Array observations}
The paper uses the observations carried out with ALMA Cycle~6 in Band-7 during 28--29 April 2019 under the 
project \#2018.1.01318.S (PI: Lokesh Kumar Dewangan). The observations were made in four spectral windows centered around 346.5 GHz, 344.3 GHz,
356.7 GHz, and 357.9 GHz, with bandwidths (and no. of channels) of 1875.0 MHz (1920), 234.0 MHz (960), 234.0 MHz (960), and 1875.0 MHz (480), respectively. 
We used the images provided by the ALMA pipeline. The source J1924-2914 was used for the flux and bandpass calibration and the source J1832-1035 was used as a phase calibrator. 
The continuum at 865 $\mu$m ($\sim$346.5 GHz) and several molecular lines have been observed toward W42-MME (see Table~\ref{tab1}). 
These data were corrected for the primary beam response. 
The images were constructed using the Briggs weighting with the robust parameter of 0.0. 
The synthesized beam size of the continuum map and the molecular line data is 0\rlap.{$''$}31 $\times$ 0\rlap.{$''$}25 (P.A. = 83$\degr$.2). 
The molecular line brightness sensitivity is achieved to be 2.4 mJy~beam$^{-1}$ {for a spectral resolution of 0.242 MHz}. 
 \begin{table*}
 % \tiny
\scriptsize
%\small
\setlength{\tabcolsep}{0.025in}
\centering
\caption{List of different spectral lines utilized in this paper.}
\label{tab1}
\begin{tabular}{lcccr}
\hline 
  Spectral window   &  Central Frequency ($\nu_{obs}$ [GHz])       & beam size \\% & rms \\   
\hline
ALMA CO(3--2)             & 345.796  & 0\rlap.{$''$}31 $\times$ 0\rlap.{$''$}25  \\
ALMA SiO(8--7)                        & 347.331  & 0\rlap.{$''$}30 $\times$ 0\rlap.{$''$}25  \\
ALMA SO 8(8)--7(7)               & 344.311  & 0\rlap.{$''$}31 $\times$ 0\rlap.{$''$}25  \\
ALMA H$^{13}$CO$^{+}$(4--3)      & 346.998  & 0\rlap.{$''$}31 $\times$ 0\rlap.{$''$}25  \\
ALMA HCO$^{+}$(4--3)             & 356.734  & 0\rlap.{$''$}30 $\times$ 0\rlap.{$''$}24  \\
ALMA CH$_{3}$OH(4$_{1,3}$--3$_{0,3}$) & 358.606  & 0\rlap.{$''$}30 $\times$ 0\rlap.{$''$}24  \\
ALMA NS(15/2--13/2)f               & 346.220  & 0\rlap.{$''$}31 $\times$ 0\rlap.{$''$}25  \\
ALMA CH$_{3}$CCH(21$_{K}$--20$_{K}$) $K$ = 0--4 &($358.709\mbox{--}818$)& 0\rlap.{$''$}30 $\times$ 0\rlap.{$''$}24  \\
ALMA CH$_{3}$CN(12$_{K}$--11$_{K}$) $K$ = 0--7 &($220.747\mbox{--}539$)& 1\rlap.{$''$}4 $\times$ 0\rlap.{$''$}8   \\
SMA CO(2--1)                 & 230.538 & 2\rlap.{$''$}2 $\times$ 1\rlap.{$''$}6  \\
SMA $^{13}$CO(2--1)                 & 220.399 & 3\rlap.{$''$}4 $\times$ 2\rlap.{$''$}4  \\
SMA C$^{18}$O(2--1)                 & 219.560 & 3\rlap.{$''$}4 $\times$ 2\rlap.{$''$}4  \\
SMA $^{13}$CS(5--4)                 & 231.221 & 2\rlap.{$''$}6 $\times$ 1\rlap.{$''$}9  \\
SMA HC$_{3}$N(24--23)               & 218.325 & 2\rlap.{$''$}6 $\times$ 2\rlap.{$''$}1  \\
SMA SiO(5--4)                       & 217.105 & 2\rlap.{$''$}6 $\times$ 2\rlap.{$''$}1  \\
SMA SO(5$_{5}$--4$_{4}$)            & 215.221 & 3\rlap.{$''$}4 $\times$ 2\rlap.{$''$}4  \\
SMA SO(5$_{6}$--4$_{5}$)            & 219.949 & 2\rlap.{$''$}6 $\times$ 2\rlap.{$''$}0  \\
SMA CH$_{3}$CN(12$_{1}$--11$_{1}$)  & 220.743 & 3\rlap.{$''$}4 $\times$ 2\rlap.{$''$}4  \\
SMA CH$_{3}$CN(12$_{0}$--11$_{0}$)  & 220.747 & 3\rlap.{$''$}4 $\times$ 2\rlap.{$''$}4  \\
SMA CH$_{3}$OH(5$_{1,4}$--4$_{2,2}$)& 216.946 & 3\rlap.{$''$}5 $\times$ 2\rlap.{$''$}5  \\
SMA H$_{2}$CO(3$_{0,3}$--2$_{0,2}$) & 218.222 & 2\rlap.{$''$}2 $\times$ 1\rlap.{$''$}7  \\
SMA H$_{2}$CO(3$_{2,2}$--2$_{2,1}$) & 218.476 & 2\rlap.{$''$}6 $\times$ 2\rlap.{$''$}0  \\
SMA H$_{2}$CO(3$_{2,1}$--2$_{2,0}$) & 218.760 & 2\rlap.{$''$}4 $\times$ 1\rlap.{$''$}9  \\
VLA CS(1--0)                  &  48.990    & 1\rlap.{$''$}7 $\times$ 1\rlap.{$''$}4  \\
\hline          
\end{tabular}
\end{table*}
\subsection{ Submillimeter Array Observations}
We carried out SMA observations (code: 2016A-A004; PI Sheng-Yuan Liu) toward W42-MME at the 230~GHz band on 2016 June 8 with the array 
in its compact configuration and on 2016 Oct with the array in its extended configuration. 

The phase center was set at the nominal position of W42-MME (i.e., $\alpha_{2000}$ = 18$^{h}$38$^{m}$14\rlap.$^{s}$54; $\delta_{2000}$ = $-$06$\degr$48$'$01\rlap.{$''$}86). 
The projected baselines range between 9~m and 86~m for the compact configuration and between 34~m and 215~m for the extended configuration.
The half-power width of the SMA primary beam is about 55$''$ at 230~GHz.
In the compact configuration, 3C273 and Titan were used as the bandpass and absolute flux calibrators, respectively.
In the extended configuration, 3C454.3 and Neptune were used as the bandpass and absolute flux calibrators, respectively.
For both occasions, nearby quasars 1733$-$130 and 1751+096 served as the complex gain calibrators.
The typical uncertainty in absolute flux density is estimated to be $\sim$20\%.

During the observing season, the array was commissioning the new SWARM (SMA Wideband Astronomical ROACH2 Machine) correlator with different LO frequencies and IF bandwidths employed in the two observing runs.
Nevertheless, the common frequency coverage extends from 215.5~GHz to 221.0~GHz in the lower side-band and from 229.5~GHz to 235.5~GHz in the upper side-band, which enables a simultaneous observation of CO(2-1), $^{13}$CO(2-1), C$^{18}$O(2-1), as well as SiO(5-4) and the CH$_3$CN $J = 12 - 11$ series. A uniform spectral resolution of 140~kHz was achieved for all channels.

We used the MIR software for data calibration and the MIRIAD package for generating the (continuum and spectral) images.
We utilized the pipeline based on the miriad-python package \citep{williams12} to process different lines simultaneously with Briggs robust=1 weighting. Due to the limited sensitivity only CO, $^{13}$CO and C$^{18}$O data were processed with the compact and the extended configuration. Other lines were restored using only the compact configuration of the array.  The SMA synthesized beam is $3{\farcs}4 \times 2{\farcs}4$, P.A. = $-$70$\degr$ for the compact data. Using the combined data, we achieved an angular resolution under robust weighting of $2{\farcs}05 \times 1{\farcs}65$, P.A. = $-$74$\degr$. The resulting molecular line brightness sensitivity is 0.1~Jy\,beam$^{-1}$ or equivalently $\sim 0.3$~K for a spectral resolution of 1.11~km\,s$^{-1}$ for a compact data, 0.2~Jy\,beam$^{-1}$ for the extended configuration, and 0.15~Jy\,beam$^{-1}$ (1.16~K) for the combined data with the same channel spacing. 
\subsection{Jansky Very Large Array Observations}
An area containing W42-MME was observed with the Jansky VLA at various epochs during 2017 February--June under program code 17A-254 (PI: Stan Kurtz). 
In February, March and May, the SiO and CS molecular lines and the
associated 7-mm narrow-band continuum were observed in the
D-configuration. 
In June, the (1,1), (2,2) and (3,3) inversion transitions of 
NH$_3$ and the associated 13-mm wide-band continuum were observed in the 
C-configuration. 
\subsubsection{CS, SiO and 7~mm continuum observations}
\label{eksec:obser}
The CS (1--0) line ($\nu_0 = 48.9909549$~GHz) and the SiO (1--0), $v=0$
line  ($\nu_0 = 43.423853$~GHz) were observed simultaneously in three 
1-hour periods, one each in February, March and May of 2017.  During each run, 
about one half-hour was spent on-source,
giving a total observing time of about 1.5 hours.  On all dates, 3C286 
was used as the flux calibrator and J1832$-$1035 was used as the phase 
calibrator. 3C286 was also used as the bandpass calibrator.
W42-MME was observed with a pointing center of $\alpha_{2000}$ = 18$^{h}$38$^{m}$14\rlap.$^{s}$54, 
$\delta_{2000}$ = $-$06$\degr$48$'$01\rlap.{$''$}86 and an assumed LSR radio velocity of +58.1~km s$^{-1}$.

Each molecular line was centered within a 32~MHz wide spectral window and observed
in dual (RR,LL) polarization mode, with 256 channels of 125~kHz each.
This provided a velocity coverage and resolution of about 200 and 0.77~km s$^{-1}$
for the CS line and  220 and 0.86~km s$^{-1}$ for the SiO line. Line-free channels
of each spectral window were used to form continuum images at 43.4 and 49.0
GHz. The beam size of the VLA 7 mm/49~GHz continuum map is $\sim$1\rlap.{$''$}7 $\times$ 1\rlap.{$''$}4, and 
the rms of this continuum map is $\sim$1.1 mJy beam$^{-1}$. 
\subsubsection{NH$_3$ and 13~mm continuum observations}
The ammonia and 13~mm ($\sim$23 GHz) continuum observations were made on five
different days in June 2017, with a 1-hour schedule block on each day.
Each day's on-source time was about 28 minutes, giving a total
integration time of about 2 hours and 20 minutes. 
The sources 3C286 and J1832$-$1035 were used as the flux and phase
calibrators, respectively; and the same pointing center (see Section~\ref{eksec:obser}) was used for W42-MME.

The three lowest ammonia inversion lines --- (1,1), (2,2) and (3,3)
--- were each observed in an 8 MHz wide spectral window comprised of
256 channels of 31.25 kHz each, thus providing a velocity coverage of
about 100~km s$^{-1}$ and a resolution of about 0.4~km s$^{-1}$ for each line.
Simultaneously, 32 spectral windows of 128 MHz each covered the
frequency range from 19 to 23 GHz with 2 MHz channels to measure
the continuum emission. The VLA 13 mm/23~GHz data were significantly affected by radio frequency interference (RFI). 
Therefore, the RFI data editing was done carefully for obtaining the final map. 
The radio continuum map is produced with Briggs weighting having a weight of $-$0.3. 
The beam size of the VLA 13 mm/23~GHz continuum map is $\sim$1\rlap.{$''$}0 $\times$ 0\rlap.{$''$}75, 
and the rms of this continuum map is $\sim$0.3 mJy beam$^{-1}$. 
\subsection{Other Archival Data}
\label{subsec:data}
We downloaded ALMA archival continuum map at 1.35 mm (resolution $\sim$1\rlap.{$''$}2 $\times$ 1\rlap.{$''$}1, P.A. = 80$\degr$.2) 
and several transitions of the CH$_{3}$CN emission toward W42-MME (see also Table~\ref{tab1}), and the target source had 12 m-array observations. We used the ALMA calibrated data in band-6 from the ALMA science archive (project \#2019.1.00195.L; PI: Molinari, Sergio), 
which were also corrected for the primary beam response. 
The observations of the project \#2019.1.00195.L were taken in four spectral windows centered around 
217.882 GHz, 218.257 GHz, 219.954 GHz, and 220.556 GHz, with bandwidths (and no. of channels) of 
1875 MHz (3840), 469 MHz (3840), 1875 MHz (3840), and 469 MHz (3840), 
respectively. The CH$_{3}$CN lines were covered in the spectral window of 220.556 GHz.

The {\it Herschel} temperature ($T_\mathrm{d}$) map \citep[resolution $\sim$12$''$;][]{molinari10b,marsh15,marsh17} 
of W42 was retrieved from the publicly available site\footnote[1]{http://www.astro.cardiff.ac.uk/research/ViaLactea/}. The SOFIA Faint Object infraRed CAmera for the SOFIA Telescope \citep[FORCAST;][]{herter12} archival images at 25.2 $\mu$m (resolution: $\sim$2\rlap.{$''$}1) and 37.1 $\mu$m (resolution: $\sim$3\rlap.{$''$}4) of W42 were downloaded from the NASA/IPAC Infrared Science Archive (Plan ID: 02\_0113; PI: James De Buizer). 
In this work, the processed level 3 data products (artifact-corrected, flux-calibrated images) were explored. 
The paper used the dust continuum map at 350 $\mu$m \citep[resolution $\sim$8\rlap.{$''$}5;][]{merello15} observed using the Second-generation Submillimeter High Angular Resolution Camera (SHARC-II) facility. The SHARC-II continuum map was exposed to a Gaussian function with a width of 3 pixels. 

We also utilized the multi-wavelength data obtained from different surveys (e.g., COHRS \citep[$^{12}$CO(J =3$-$2); resolution $\sim$16$''$; rms $\sim$1 K;][]{dempsey13}, 
{\it Herschel} Infrared Galactic Plane Survey \citep[Hi-GAL; $\lambda$ = 70--500 $\mu$m; resolution $\sim$5\rlap.{$''$}.8--37$''$;][]{molinari10}, 
and Galactic Legacy Infrared Mid-Plane Survey Extraordinaire \citep[GLIMPSE: 3.6--8.0 $\mu$m; resolution $\sim$2$''$;][]{benjamin03}). 
This work also used the published continuum-subtracted H$_{2}$ image (resolution $\sim$0\rlap.{$''$}8) and the VLT/NACO adaptive-optics images 
at K$_{s}$-band and L$^{\prime}$-band (resolution $\sim$0\rlap.{$''$}1--0\rlap.{$''$}2), which were taken from \citet{dewangan15b}. The K-band polarimetric data were obtained from \citet{jones04}. Additionally, we obtained the GPS~6~cm Epoch~3 radio continuum map (beam size $\sim$2$''$ $\times$ 1\rlap.{$''$}6) downloadable from the MAGPIS website\footnote[2]{http://third.ucllnl.org/gps}. The GPS radio continuum data were observed with the VLA B-configuration (2006).  
\section{Results}
\label{sec:data}
\subsection{Continuum Emission}
\label{sec:conta}
In this section, we present the infrared, sub-millimeter (sub-mm), millimeter (mm), and 
centimeter (cm) continuum images of W42-MME. 
\subsubsection{mm and cm continuum maps}
\label{sec:conta1x}
\begin{figure*}
\epsscale{1.1}
\plotone{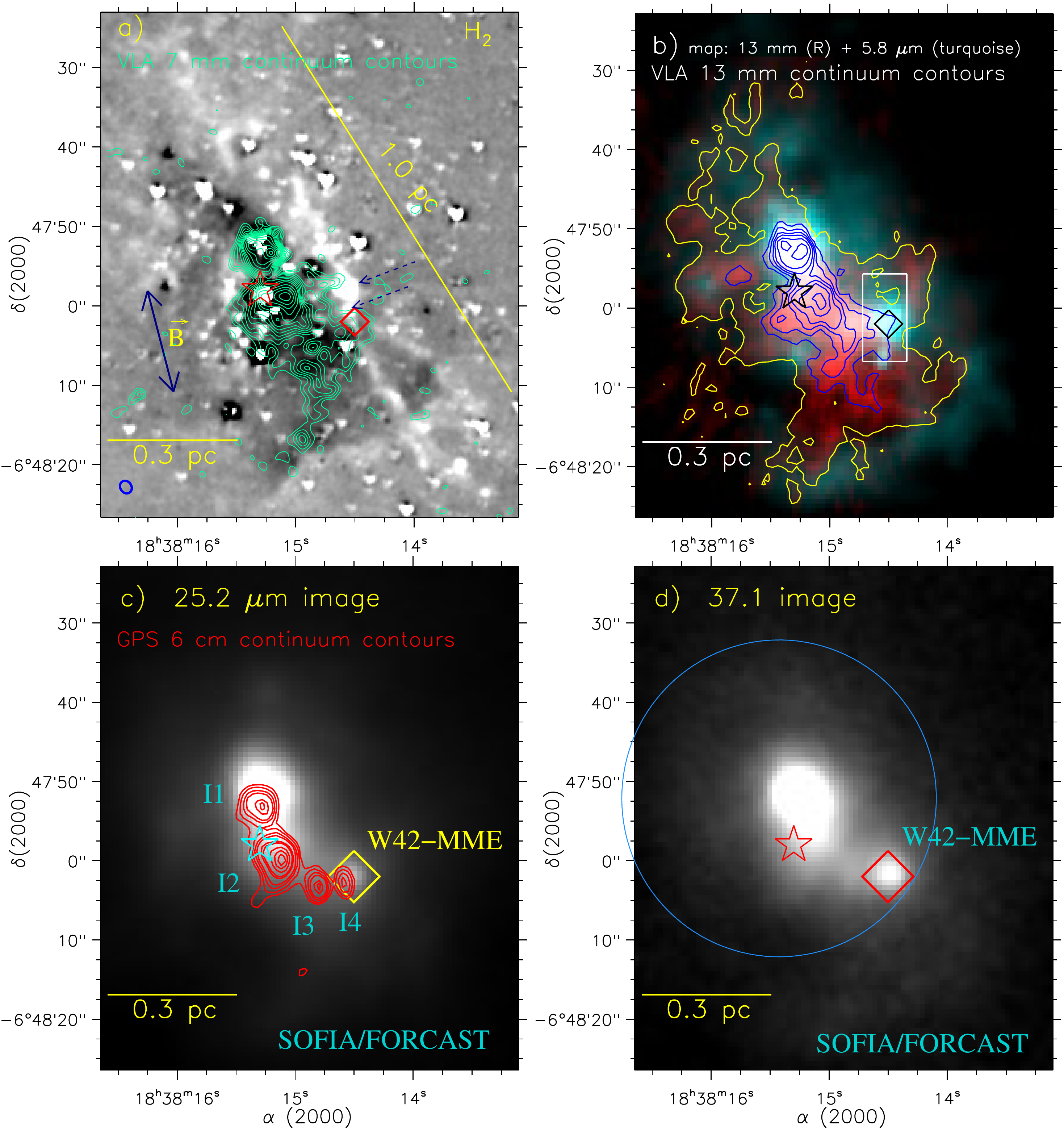}
\caption{
a) Overlay of the VLA 7 mm (or 49~GHz) radio continuum emission contours on the continuum-subtracted H$_\mathrm{2}$ image. The synthesized beam of the VLA 7 mm is 1\rlap.{$''$}7 $\times$ 1\rlap.{$''$}4, P.A. = 34$\degr$.8 (lower left corner). 
The contour levels of the continuum emission are at (0.1, 0.15, 0.2, 0.25, 0.3, 0.35, 0.4, 0.45, 0.5, 0.6, 0.7, 0.8, 0.9, 0.98) $\times$ 30.76 mJy beam$^{-1}$ (1$\sigma$ $\sim$1.1 mJy beam$^{-1}$). 
The magnetic field (B) direction \citep[taken from][]{jones04} is indicated by a thick blue arrow. b) The panel displays a two color-composite map (VLA 13 mm (red) and {\it Spitzer} 5.8 $\mu$m (turquoise) images) of W42. 
The color-composite map is also overlaid with the VLA 13 mm (or 23~GHz) radio continuum emission contours (in blue and yellow; beam size $\sim$1\rlap.{$''$}0 $\times$ 0\rlap.{$''$}75). The yellow contour is plotted at 1.27 mJy beam$^{-1}$, while the blue contours are shown with the levels of 3.38, 5.06, 6.75, 8.44, 10.97, and 
14.35 mJy beam$^{-1}$ (1$\sigma$ $\sim$0.3 mJy beam$^{-1}$). 
The solid box (in white) encompasses the area shown in Figures~\ref{fig2}c,~\ref{fig2}d, and~\ref{fig2}e. 
c) Overlay of the GPS 6~cm Epoch~3 radio continuum contours (beam size $\sim$2$''$ $\times$ 1\rlap.{$''$}6) 
on the SOFIA/FORCAST image at 25.2 $\mu$m. The contour levels are 2.3, 3, 4, 5, 7.5, 8.5, 11.5, 14, and 16 
mJy beam$^{-1}$ (1$\sigma$ $\sim$0.54 mJy beam$^{-1}$). 
d) The panel shows the SOFIA/FORCAST image at 37.1 $\mu$m. A big circle shows an area presented in Figure~\ref{fig2l}a. 
In each panel, the positions of a 6.7 GHz MME (diamond symbol) and an O5--O6 star (star symbol) are marked. 
In all panels, the scale bar shows a size of 0.3 pc at a distance of 3.8~kpc.}  
\label{fig1}
\end{figure*}

In Figure~\ref{fig1}a, we show the overlay of the VLA 7 mm/49~GHz radio continuum emission 
contours on the H$_\mathrm{2}$ image, showing the wide-scale environment of W42-MME. 
In the background map, the parsec scale H$_\mathrm{2}$ outflow with H$_\mathrm{2}$ knots/bow-shock feature is evident. 
In the immediate vicinity of the 6.7 GHz MME, at least two H$_\mathrm{2}$ knots are indicated by arrows. 
Note that the continuum and line data (from VLA/SMA/ALMA) used in this paper do not cover the area containing the entire parsec scale H$_\mathrm{2}$ outflow.

In Figure~\ref{fig1}b, we overlay the VLA 13 mm/23~GHz radio continuum emission contours on a two color-composite map (VLA 13 mm (red) and {\it Spitzer} 5.8 $\mu$m (turquoise) images) of W42. 
The VLA 13 mm continuum emission is distributed well within an emission structure seen in the 
{\it Spitzer} 5.8 $\mu$m image. Earlier, this structure was reported as an ionized cavity-like 
feature using the H$_{2}$ emission, the {\it Spitzer} 5.8 $\mu$m image, and the radio continuum emission \citep[see Figure~2a in][]{dewangan15b}.

Figures~\ref{fig1}c and~\ref{fig1}d show the SOFIA 25.2 $\mu$m and 37.1 $\mu$m 
continuum images, respectively. 
We also observe the MYSO W42-MME in both the SOFIA infrared images. 
The observed NIR emission from the protostars is mainly due to scattered light escaping 
from the cavities \citep[e.g.,][]{zhang11}, whereas the observed mid-infrared emission can be 
explained as thermal emission from warm dust in the outflow cavity walls. 
In Figure~\ref{fig1}c, we also display the GPS 6~cm Epoch~3 radio continuum contours overlaid on 
the SOFIA continuum image at 25.2 $\mu$m. 
Using the {\it clumpfind} IDL program \citep{williams94}, four ionized clumps are identified 
in the GPS 6~cm continuum map, and are labeled as I1, I2, I3, and I4 in Figure~\ref{fig1}c. 
In addition to the total flux, the {\it clumpfind} also gives the full width at half maximum (FWHM) 
not corrected for beam size for the x-axis (i.e., FWHM$_{x}$), and for the y-axis (i.e., FWHM$_{y}$). 
The total fluxes (FWHM$_{x}$ $\times$ FWHM$_{y}$) of the ionized clumps I1, I2, I3, and I4 are about 32.7 mJy (3\rlap.{$''$}3 $\times$ 2\rlap.{$''$}9), 74.4 mJy (3\rlap.{$''$}9 $\times$ 3\rlap.{$''$}0), 17.8 mJy (2\rlap.{$''$}6 $\times$ 1\rlap.{$''$}9), and 7.9 mJy (2\rlap.{$''$}2 $\times$ 1\rlap.{$''$}4) mJy, respectively. 
We also computed the total fluxes of I1(I2) at 13 mm and 7 mm to be $\sim$300($\sim$400) mJy and $\sim$170($\sim$270) mJy, respectively 
(see Figures~\ref{fig1}a and~\ref{fig1}b). On the basis of the spectral index calculation between 6 cm and 13 mm, 
the sources I1 and I2 are optically thick at 6 cm. 

On the other hand, no emission is detected toward the ionized clumps I3 and I4 in the maps at 13 and 7 mm. 
Hence, we estimated upper limits on fluxes of I3(I4) to be of $\sim$22($\sim$17) mJy and $\sim$17($\sim$15) mJy at 13 mm and 7 mm, respectively. These estimates take into account inhomogeneities of the diffuse emission observed toward the area hosting these sources. 
Concerning the sources I3 and I4, these estimates are in agreement with optically thin emission from 6 cm to 7 mm. 
Following \citet{dewangan21}, we compute the number of Lyman continuum photons $N_\mathrm{UV}$ of only two ionized clumps, I3 and I4 \citep[see also][]{matsakis76}, which seem to be optically thin at 6 cm. The calculation uses the observed flux values, electron temperature = 10000~K and distance = 3.8 kpc. 
We determined $\log{N_\mathrm{UV}}$ of I3 and I4 to be 46.4 and 46.0 s$^{-1}$, respectively. 
Using the reference of \citet{panagia73}, both these ionized clumps are powered by a massive B-type star.
Note that the ionized clump I4 is traced near the position of W42-MME, which has an offset of about 1\rlap.{$''$}4. 
However, the VLA 7 mm and 13 mm continuum maps do not show any radio counterpart of W42-MME. 

Noticeable radio continuum emission is observed around 
a region containing the position of the O5-O6 star in the observed 6~cm, 7 mm, and 13 mm continuum maps (see the ionized clumps I1 and I2), where the infrared emission is also detected in the SOFIA images at 25.2 and 37.1 $\mu$m (see Figure~\ref{fig1}). 
As per our calculations, we find that the sources I1 and I2 are optically thin at 13 mm (23 GHz). 
Hence, using the observed fluxes at 13 mm, we estimate $\log{N_\mathrm{UV}}$ of the ionized clumps I1 and I2 
to be 47.6 and 47.8 s$^{-1}$, respectively. Following \citet{panagia73}, both these ionized clumps are excited by a massive B0V-O9.5V star. However, a detailed study of I1 and I2 is beyond the scope of this paper. This work mainly focuses around W42-MME.  

\begin{figure*}
\epsscale{0.9}
\plotone{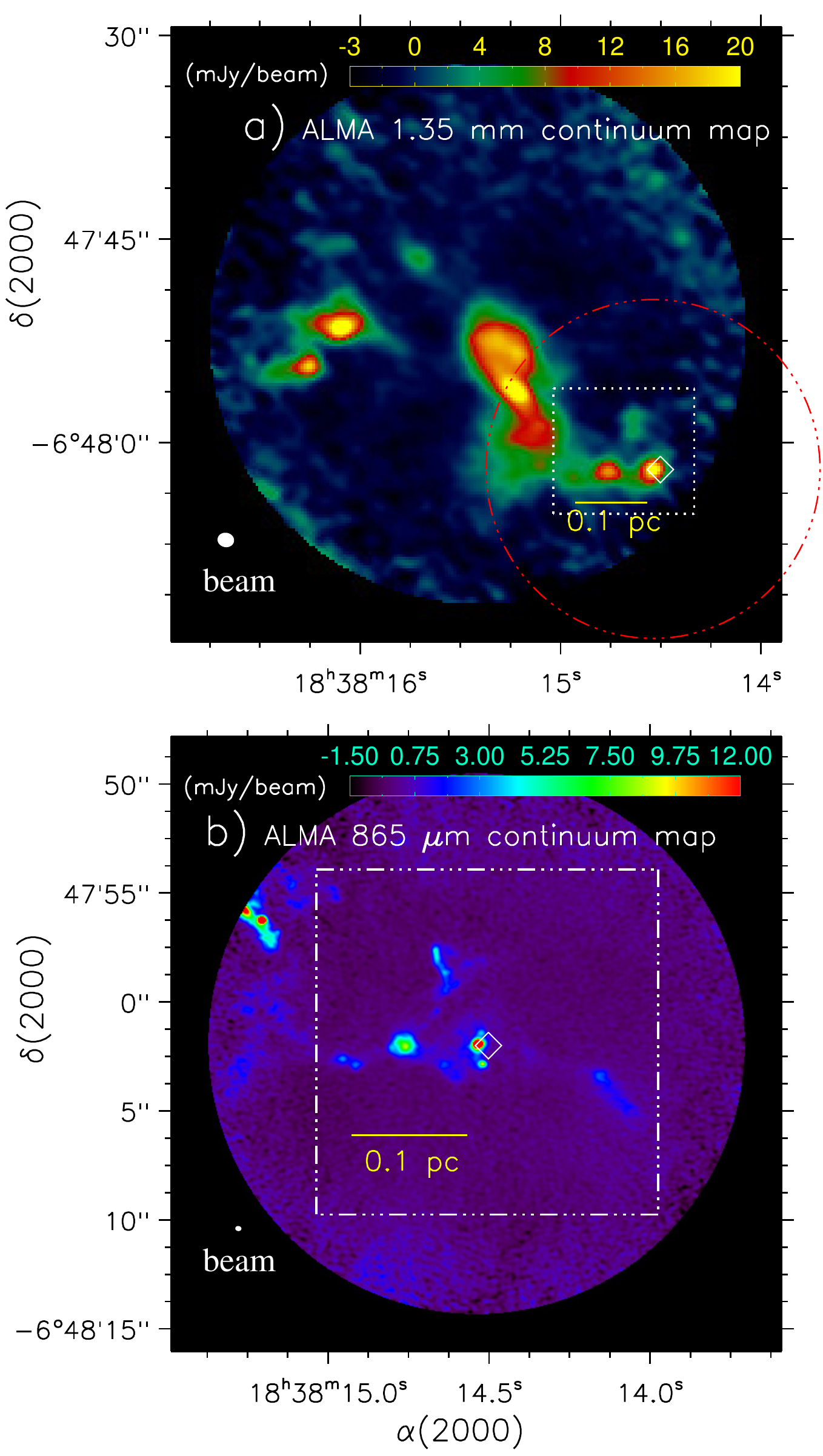}
\caption{a) The panel displays the 1.35 mm continuum map in the ALMA band-6. 
The synthesized beam is 1\rlap.{$''$}2 $\times$ 1\rlap.{$''$}1, P.A. = 80$\degr$.2 (lower left corner). 
A big dotted-dashed circle highlights an area shown in Figure~\ref{fig2l}b, while 
a dotted box shows an area presented in Figure~\ref{fig2}a. 
b) The panel displays a zoomed-in area as shown in Figure~\ref{fig2l}a (red circle) using the 865 $\mu$m continuum image in the ALMA band-7. The synthesized beam is 0\rlap.{$''$}29 $\times$ 0\rlap.{$''$}23, P.A. = 83$\degr$.2 (lower left corner). 
A dotted-dashed box encompasses an area presented in Figures~\ref{fig3}a and~\ref{fig3}c. 
In each panel, the position of a 6.7 GHz MME is marked by a diamond symbol.} 
\label{fig2l}
\end{figure*}

\begin{figure*}
\epsscale{1.1}
\plotone{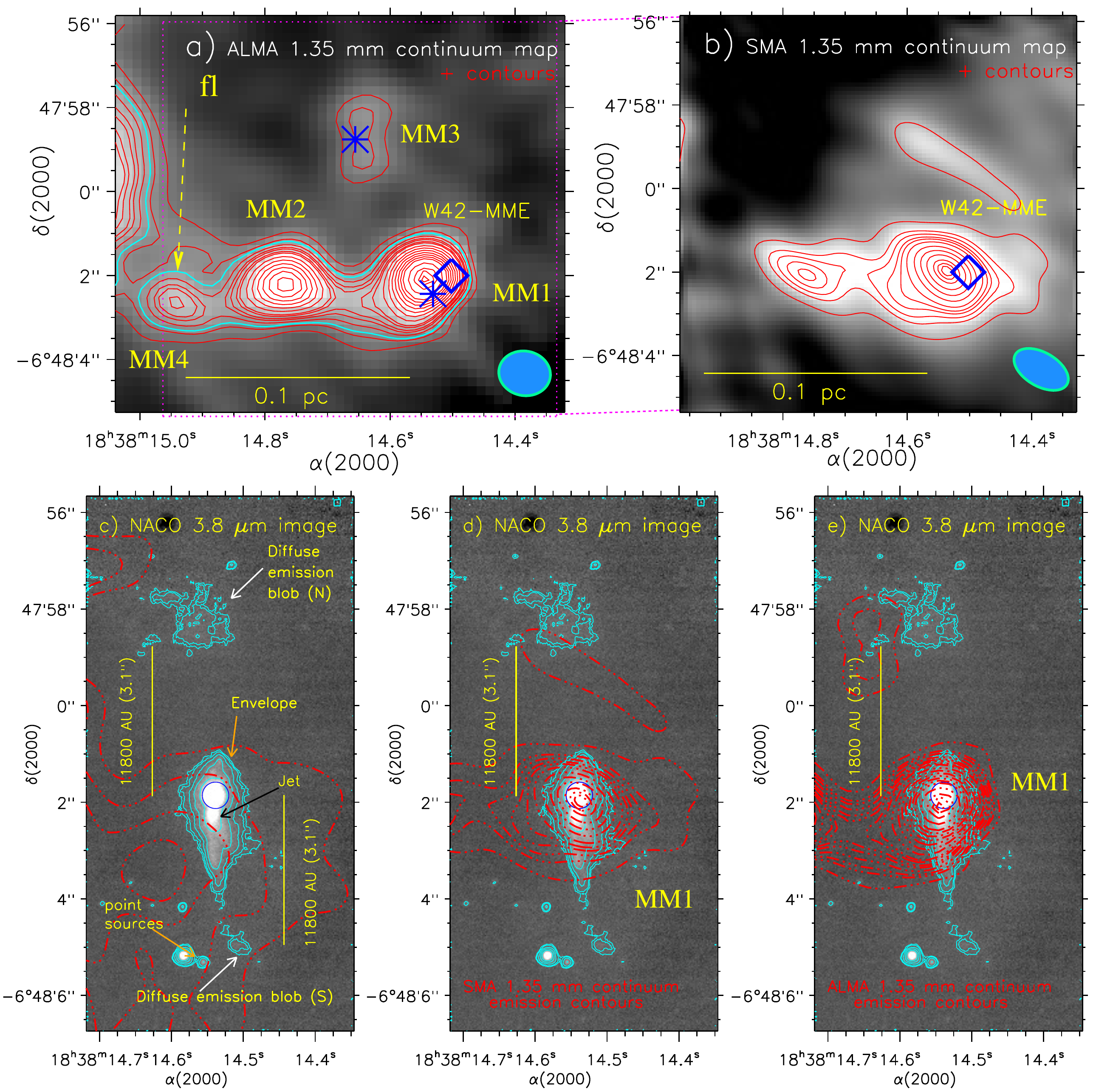}
\caption{a) The panel presents the ALMA continuum map at 1.35 mm. 
The continuum emission contours (in red) are plotted at (0.15, 0.2, 0.22, 0.25, 0.28, 0.3, 0.33, 0.35, 0.4, 0.45, 0.5, 
0.55, 0.6, 0.65, 0.7, 0.8, 0.9, 0.95, 0.98) $\times$ 23.4 mJy beam$^{-1}$ (1$\sigma$ $\sim$0.55 mJy beam$^{-1}$). 
An elongated feature is highlighted by a solid contour (in cyan) at 5.153 mJy beam$^{-1}$. 
A dotted box (in magenta) highlights an area shown in Figure~\ref{fig2}b. 
The positions of water masers are indicated by asterisks \citep[from][]{walsh14}. 
b) The panel displays the SMA 1.35 mm continuum map. The continuum emission contours (in red) are shown 
at (0.2, 0.3, 0.35, 0.4, 0.44, 0.5, 0.6, 0.7, 0.8, 0.9, 0.95, 0.98) 
$\times$ 44.7 mJy beam$^{-1}$ (1$\sigma$ $\sim$2.2 mJy beam$^{-1}$). 
The synthesized beam is 1\rlap.{$''$}4 $\times$ 0\rlap.{$''$}8, P.A. = 60$\degr$.2 (lower right corner). 
c) Overlay of the VLA 7 mm radio continuum emission on the VLT/NACO adaptive-optics L$'$ image ($\lambda$ = 3.8 $\mu$m; resolution $\sim$0\rlap.{$''$}1) around W42-MME. The VLA 7 mm radio continuum emission is shown by dotted-dashed red contours, which are the same as 
in Figure~\ref{fig1}a.  
The NACO image is also overlaid with the L$'$ contours (in cyan). 
d) Overlay of the SMA continuum emission contours at 1.35 mm on the VLT/NACO L$'$ image.  
The dotted-dashed red contours are the same as in Figure~\ref{fig2}b. 
e) Overlay of the ALMA continuum emission contours at 1.35 mm on the VLT/NACO L$'$ image. 
The dotted-dashed red contours are the same as in Figure~\ref{fig2}a. 
In panels ``a" and ``b", the position of a 6.7 GHz MME is marked by a diamond symbol. 
In panels ``d" and ``e", the background map is similar to the one shown in Figure~\ref{fig2}c.} 
\label{fig2}
\end{figure*}

Figure~\ref{fig2l}a displays the ALMA band-6 (1.35 mm) continuum 
map (beam size $\sim$1\rlap.{$''$}2 $\times$ 1\rlap.{$''$}1) of W42, and the field of view of this observation is indicated by a big circle (radius $\sim$20$''$) in Figure~\ref{fig1}d. In Figure~\ref{fig2l}b, we present the ALMA band-7 (865 $\mu$m) 
continuum map (beam size $\sim$0\rlap.{$''$}29 $\times$ 0\rlap.{$''$}23 or 1100 AU $\times$ 875 AU) of W42-MME. 
A dotted-dashed circle (radius $\sim$12\rlap.{$''$}5) outlines the field of view of the band-7 observation in Figure~\ref{fig2l}a. 
A detailed examination of the ALMA band-7 continuum map is presented in Section~\ref{sec:conta2x}. 
Figures~\ref{fig2}a and~\ref{fig2}b display zoomed-in views of the ALMA band-6 and SMA continuum maps at 1.35 mm of W42-MME, respectively. 
Both the maps have a beam size of $\sim$1$''$.
The ALMA band-6 continuum map reveals three compact mm sources (i.e., MM1, MM2, and MM4) as well as one more mm continuum source MM3, which does not appear as a compact object. The three continuum sources MM1, MM2, and MM4 appear to coincide with a diffuse, 
elongated feature (``fl"; extent $\sim$0.15 pc) in the ALMA band-6 continuum map (see the cyan contour at 5.153 mJy beam$^{-1}$ in 
Figure~\ref{fig2}a). The two sources (i.e., MM1 and MM2) are spatially seen over a scale of 0.1 pc in the SMA continuum map, and are found near the ionized clumps (i.e., I4 and I3; see Figure~\ref{fig1}c). The other two continuum sources 
(i.e., MM3 and MM4) do not coincide with the radio continuum peaks traced in the GPS continuum map at 6~cm.

In Figures~\ref{fig2}c,~\ref{fig2}d, and~\ref{fig2}e, we present the VLT/NACO L$'$ image overlaid with 
the continuum emission contours at VLA 7 mm, SMA 1.35 mm, and ALMA 1.35 mm, respectively. 
In these figures, the contours of L$'$ image are also presented to display the diffuse infrared emission, 
the proposed infrared jet, and the outflow cavity/infrared envelope. 
A small circle is also marked to highlight the position of the powering 
source of the infrared outflow/jet. 
The jet-outflow W42-MME system is embedded within the compact source MM1. The mm continuum source MM3 appears to be associated with the diffuse infrared emission traced in the L$'$ image, where the H$_{2}$ emission is also detected (see an arrow in Figure~\ref{fig1}b). Furthermore, the position of a water maser \citep[from][]{walsh14} is also observed toward the sources MM1 and MM3 (see asterisks in Figure~\ref{fig2}a).
 
\subsubsection{Determination of the mass of mm continuum sources}
\label{sec:conta1xxx}
The mass of each compact continuum source is estimated with the knowledge of its integrated flux and temperature. 
In this relation, we employed the {\it clumpfind} IDL program to compute the integrated fluxes at ALMA 1.35 mm of the continuum sources. We examined the {\it Herschel} temperature ($T_\mathrm{d}$) map to obtain the dust temperature 
toward W42-MME, which is found to be $\sim$40~K. 
The mass of the mm continuum source was computed using the following formula \citep{hildebrand83}:
\begin{equation}
M \, = \, \frac{D^2 \, S_\nu \, R_t}{B_\nu(T_D) \, \kappa_\nu}
\end{equation} 
\noindent where $S_\nu$ is the integrated 1.35 mm flux (in Jy), 
$D$ is the distance (in kpc), $R_t$ is the gas-to-dust mass ratio (assumed to be 100), 
$B_\nu$ is the Planck function for a dust temperature $T_D$, 
and $\kappa_\nu$ is the dust absorption coefficient. Here, we used $\kappa_\nu$ = 0.9\,cm$^2$\,g$^{-1}$ at 1.3 mm \citep{ossenkopf94},  
$D$ = 3.8 kpc, and $T_D$ = 40 K. 
The flux densities (FWHM$_{x}$ $\times$ FWHM$_{y}$) of the compact continuum sources MM1, MM2, and MM4 in continuum are 
about 36.0 mJy (1\rlap.{$''$}49 $\times$ 1\rlap.{$''$}23), 26.6 mJy (1\rlap.{$''$}59 $\times$ 0\rlap.{$''$}99), and 7.7 mJy (1\rlap.{$''$}25 $\times$ 0\rlap.{$''$}63), respectively. 
The masses of MM1, MM2, and MM4 are estimated to be $\sim$5.2, $\sim$3.9, and $\sim$1.1 M$_{\odot}$, respectively. 
If we use $T_D$ = 70 K (see Section~\ref{frottemp1a}) then the masses of MM1, MM2, and MM4 are computed to be $\sim$2.8, $\sim$2.1, and $\sim$0.6 M$_{\odot}$, respectively. As mentioned earlier, the ionized clump I4 overlaps with the source MM1 at 1.35 mm.   
Hence, free-free emission from I4 can contribute to the mm flux of MM1. We computed this contribution from the 6 cm flux assuming the spectral index of $-$0.1 typical for optically thin free-free emission. In this way, we obtain the upper limit for this contribution of 5.4 mJy, which is about 15\% of the derived flux value of 36.0 mJy. On the basis of the corrected flux value (i.e., 30.6 mJy), the mass of MM1 is determined to be $\sim$4.4(2.4) M$_{\odot}$ at $T_D$ = 40(70)~K. In the case of MM2 and MM4, we are unable to estimate a possible free-free contribution.  
In general, the estimation of the mass suffers from various uncertainties, which include the assumed dust temperature, opacity, 
and measured flux. Hence, the uncertainty in the mass estimate of each continuum source could be typically $\sim$20\% and 
at largest $\sim$50\%. 

We also computed the total mass of the elongated feature ``fl" (extent $\sim$0.15 pc) 
to be $\sim$11.2 M$_{\odot}$ at $T_D$ = 40 K. 
This estimation uses the integrated flux of the elongated feature (i.e., $\sim$77.4 mJy), 
which was determined using the {\it clumpfind} IDL program. 
If we treat this feature as a filament having a high aspect ratio (length/diameter) then its line mass, or mass per unit length (i.e., M$_\mathrm{line,obs}$) is determined to be $\sim$75 M$_{\odot}$ pc$^{-1}$. 
One can define a critical line mass M$_{\rm line,crit}$ for a gas
filament, modeled as an infinitely long, self-gravitating, isothermal
cylinder without magnetic support, given by M$_{\rm line,crit} \sim 16$
M$_\odot$~pc$^{-1} \times ({\rm T}_{gas}/10~{\rm K})$ \citep[e.g.,][]{ostriker64,inutsuka97,andre14}. 
Our value of $\sim$75 M$_{\odot}$ pc$^{-1}$ suggests
that the feature ``fl" is a thermally supercritical filament.
It is thought that thermally supercritical filaments are prone 
to radial gravitational collapse and fragmentation \citep[e.g.,][]{andre10}. 
In general, a factor of ``{$\cos i$}" is involved in the estimate of the line mass of a filament \citep[i.e.,][]{kainulainen16}, where "{\it i}" is the angle between the sky plane and the filament's major axis. Here, we consider the filament lying in the sky plane, resulting in ``{$\cos i$} = 1".
\subsubsection{ALMA sub-mm continuum map}
\label{sec:conta2x}
The continuum emission of MM1 at 1.35 mm may arise from the dusty envelope and disc surrounding the MYSO W42-MME. 
However, the resolution of the ALMA band-6 and SMA 
continuum maps at 1.35 mm ($\sim$1$''$) is not enough to further resolve the disk or the inner circumstellar substructures of MM1 
(see Figures~\ref{fig2l}a and~\ref{fig2l}b). 

\begin{figure*}
\epsscale{1}
\plotone{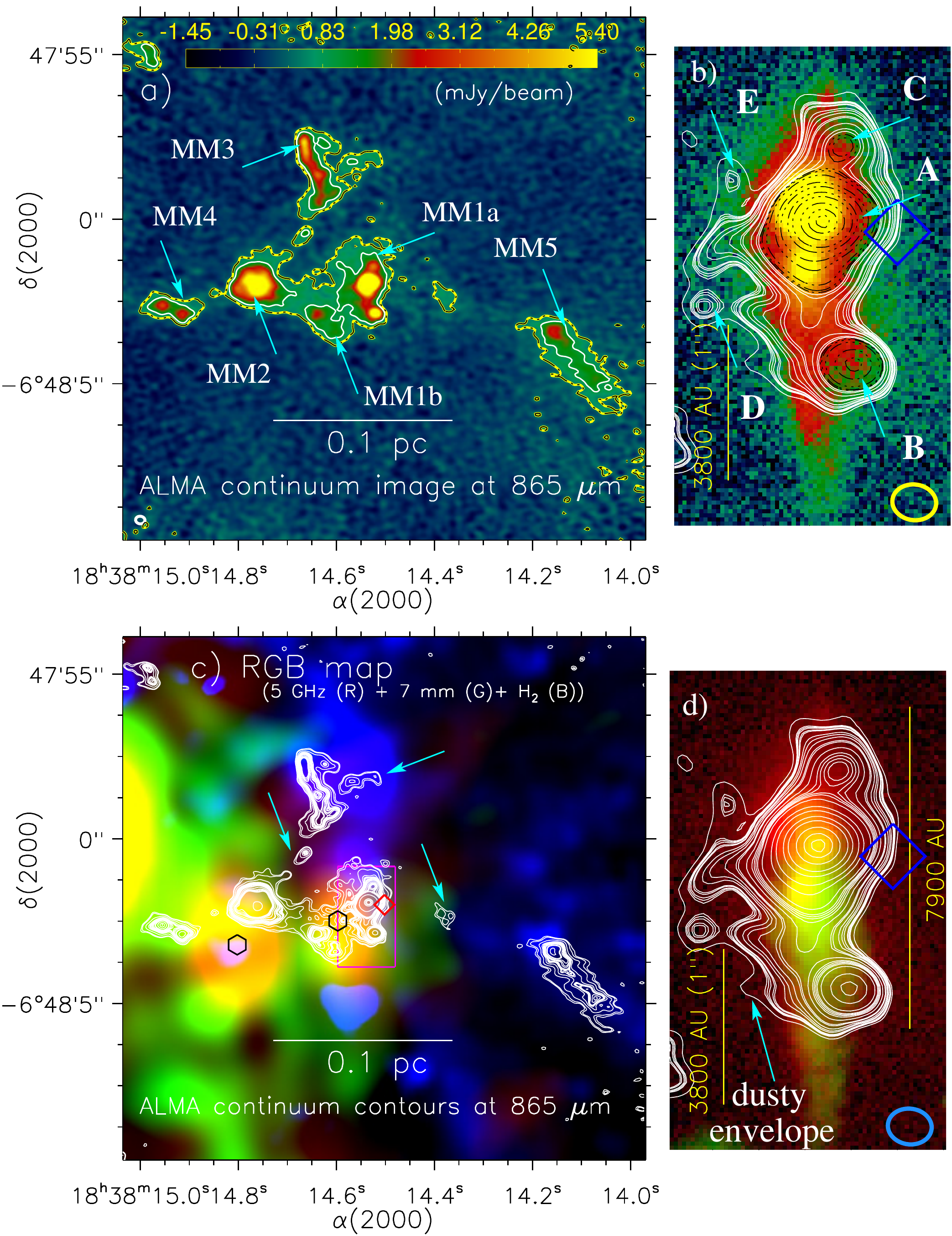}
\caption{a) The panel displays the 865 $\mu$m continuum image in the ALMA band-7. 
The synthesized beam is 0\rlap.{$''$}29 $\times$ 0\rlap.{$''$}23, P.A. = 83$\degr$.2 (lower left corner). 
The ALMA continuum emission contour (in white) at 865 $\mu$m is also overlaid with a level of 1.1 mJy beam$^{-1}$. 
A broken contour (in yellow and black; at 0.45 mJy beam$^{-1}$) is also shown to highlight extended features seen in the continuum map. Six cores (i.e., MM1a, MM1b, MM2--5) are labeled in the map. 
b) A zoomed-in view of the NACO L$'$ image around MM1a (see a solid box in Figure~\ref{fig3}c). 
The L$'$ image is overlaid with the ALMA continuum emission at 865 $\mu$m. 
The contours are at (0.029, 0.0325, 0.033, 0.0335, 0.0345, 0.035, 0.038, 0.04, 0.043, 0.048, 0.05, 0.06, 0.07, 0.075, 0.08, 0.085, 0.1, 0.15, 0.2, 0.3, 0.4, 0.6, 0.8, 0.95) $\times$ 42.59 mJy beam$^{-1}$ (1$\sigma$ $\sim$0.15 mJy beam$^{-1}$). Five continuum sources (i.e., A--E) are also labeled in the map. 
c) Overlay of the ALMA continuum emission contours at 865 $\mu$m on a three color-composite map (GPS 5 GHz (red), 7 mm (green), and 
H$_{2}$ (blue) images). The contour levels are at (0.0125, 0.015, 0.022, 0.029, 0.0325, 0.033, 0.0335, 0.0345, 0.035, 0.038, 0.04, 0.043, 0.048, 0.05, 0.06, 0.07, 0.075, 0.08, 0.085, 0.18, 0.65, 0.80, 0.90) $\times$ 42.59 mJy beam$^{-1}$ (1$\sigma$ $\sim$0.15 mJy beam$^{-1}$). 
A solid box (in magenta) indicates an area presented in Figures~\ref{fig3}b and~\ref{fig3}d. 
Black hexagons show the peak positions of the radio continuum sources seen in the GPS 5 GHz continuum map. 
d) A zoomed-in view of a two-color composite NACO map (L$'$ (red) and K$_{s}$ (green) images) around MM1a (see a solid box in Figure~\ref{fig3}c). 
The panel displays the ALMA continuum emission contours at 865 $\mu$m as shown in Figure~\ref{fig3}b. 
In panels ``b--d", the position of a 6.7 GHz MME is marked by a diamond symbol. 
} 
\label{fig3}
\end{figure*}

In Figure~\ref{fig3}a, we present the ALMA continuum map at 865 $\mu$m, where a broken contour at 0.45 mJy~beam$^{-1}$ (3$\sigma$) is drawn to indicate the extent of the continuum emission. 
The 865 $\mu$m continuum contour (in white) at 1.1 mJy~beam$^{-1}$ is also overlaid on the map, tracing the six distinct cores (MM1a, MM1b, MM2--5). 
The source MM1a is seen toward MM1, while the source MM1b is detected between MM1 and MM2. 
Note that the core MM5 is located outside the 1.35 mm map area of Figures~\ref{fig2}a--\ref{fig2}b. 
The 865 $\mu$m color scale shows that at the higher frequency and resolution of the ALMA band 7 observations, 
the mm cores lying along the filament have internal structure with multiple components. 

Figure~\ref{fig3}c shows a three color-composite map (GPS 5 GHz/6 cm (red), 7 mm (green), and 
H$_{2}$ (blue) images) overlaid with the ALMA continuum emission contours at 865 $\mu$m. 
The color-composite map shows that none of the mm sources MM1--5 coincide with the radio continuum peaks (see black hexagons). 
Arrows indicate three additional continuum sources that were not detected at 1.35 mm. 

The NACO L$'$ image around MM1a is presented in Figure~\ref{fig3}b, while a two-color composite NACO image (L$'$(red) + K$_{s}$ (green)) around MM1a is displayed in Figure~\ref{fig3}d. The NACO color-composite map is taken from \citet{dewangan15b}. 
The position of the 6.7 GHz MME is indicated by a diamond symbol in Figures~\ref{fig3}b--\ref{fig3}d. 
The infrared envelope/outflow cavity, the powering source of the H$_{2}$ outflow, and the proposed infrared jet are brighter in the L$^{\prime}$ image than the K$_{s}$ image. 
The diffuse emission observed in the K$_{s}$ image coincides well with the jet-like feature, which has been suggested 
as an ionized jet \citep[see][for more details]{dewangan15b}. 

In Figures~\ref{fig3}b and~\ref{fig3}d, the ALMA continuum emission contours at 865 $\mu$m are also shown, 
revealing the substructure of the western part of core MM1. 
The inner circumstellar structure of W42-MME traced in the NACO images and the ALMA continuum map can be 
compared in these figures, showing a dusty or circumstellar envelope (extent $\sim$7900 AU) surrounding the MYSO W42-MME 
(see an outer contour in Figures~\ref{fig3}b and~\ref{fig3}d). 
The ALMA continuum map further reveals five continuum peaks ``A--E" inside the dusty envelope. 
Two of these, ``A" and ``B", are resolved by the ALMA beam. 
Furthermore, the previously reported infrared envelope/outflow cavity is also 
seen toward the dusty envelope. 
The peak flux density (Rayleigh-Jeans temperature or radiation temperature) of A, B, C, D, and E is found to be 
4.36 mJy beam$^{-1}$ (6.5 K), 0.94 mJy beam$^{-1}$ (1.4 K), 0.39 mJy beam$^{-1}$ (0.58 K), 
0.15 mJy beam$^{-1}$ (0.23 K), and 0.14 mJy beam$^{-1}$ (0.21 K), respectively. 
Here, a conversion factor between flux density and radiation temperature for the ALMA beam is $\sim$150 K Jy$^{-1}$ \citep[see also][]{zinchenko20}. 
The ALMA continuum source ``A" hosts the powering source of the H$_{2}$ outflow and the proposed infrared jet. 
The H$_{2}$ knots are seen to the north of 
source ``A" (Figure~\ref{fig3}c). The other continuum peaks/sources ``B--E" are 
in the immediate surroundings of the continuum source ``A" within the dusty envelope. 

The detection of MM1 in the maps at 1.35 mm and 865 $\mu$m enables us to estimate 
its spectral index, which is found to be about 3.6, favoring an optically thin dust emission. 
This calculation includes the correction for the I4 contribution. 
In the case of MM2, the spectral index between 1.35 mm and 865 $\mu$m is determined to be $\sim$3. 
Mass estimates from the 865 $\mu$m continuum map should be more reliable because the relative contribution 
of the free-free emission at this wavelength is much lower than at 1.35 mm.

Using Equation~1, we also computed the masses of six continuum sources (i.e., MM1a, MM1b, MM2, MM3, MM4, and MM5) detected in the ALMA map at 865\,$\mu$m. 
It is noted that the sources ``A--E" are seen in the direction of the continuum source MM1a. 
Table~\ref{tab2} contains physical parameters (i.e., position, flux density, FWHM$_{x}$ $\times$ FWHM$_{y}$, and mass) of 
six continuum sources. 
The integrated fluxes at 865 $\mu$m were obtained using the {\it clumpfind} IDL program. 
In the calculations, we used $\kappa_\nu$ = 1.85\,cm$^2$\,g$^{-1}$ at 865\,$\mu$m \citep{schuller09}, $D$ = 3.8 kpc, 
and $T_D$ = [40, 70]~K. 
The mass of the source ``A" is also estimated to be $\sim$2.2 and $\sim$1.2 M$_{\odot}$ at $T_D$ = 40 and 70 K, respectively. 
Here, the flux density (FWHM$_{x}$ $\times$ FWHM$_{y}$) of the continuum source ``A" is 
determined to be $\sim$71.8 mJy (0\rlap.{$''$}35 $\times$ 0\rlap.{$''$}35). 
\subsection{Molecular Line Emission}
\label{sec:line}
Several molecular lines are detected toward our target source, and are observed by the SMA, VLA, ALMA band-6 and ALMA band-7 facilities. 
In the direction of W42-MME, the molecular emission is mainly studied in a velocity range of [60, 70] km s$^{-1}$. 
\subsubsection{Molecular Lines from the SMA and VLA}
\begin{figure*}
\epsscale{1}
\plotone{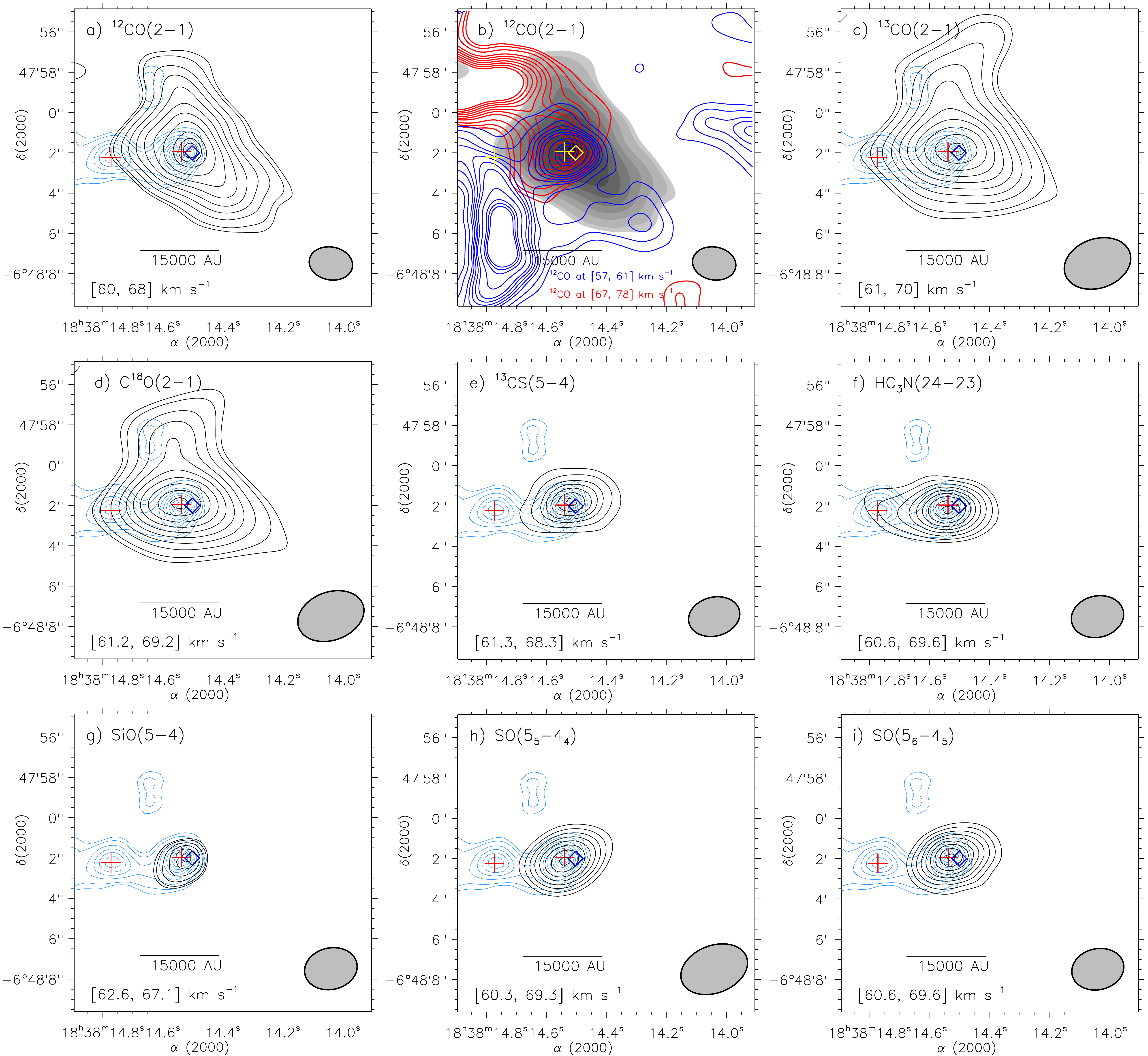}
\caption{Panels show the SMA molecular line detections.  
a) Contour map of the CO(2--1) integrated intensity emission at [60, 68] km s$^{-1}$. 
The contour levels are 
at (0.16, 0.2, 0.3, 0.4, 0.5, 0.6, 0.7, 0.8, 0.9, 0.95, 0.98) $\times$ 66.64 Jy beam$^{-1}$ km s$^{-1}$ 
(1$\sigma$ $\sim$2 Jy beam$^{-1}$ km s$^{-1}$). 
b) Overlay of the CO(2--1) emission contours (at [57, 61] and [67, 78] km s$^{-1}$) on 
the CO(2--1) integrated intensity filled contour map (see Figure~\ref{fig4}a). 
The CO(2--1) emission from 57 to 61 km s$^{-1}$ is shown by solid blue contours (i.e., blueshifted component), 
and the contour levels are at 0.42, 0.83, 1.25, 1.46, 1.67, 1.88, 2.08, 2.50, 2.92, 3.13, and 3.29 
Jy beam$^{-1}$ km s$^{-1}$. 
The CO(2--1) emission from 67 to 78 km s$^{-1}$ is drawn by solid red contours (i.e., redshifted component), and 
the contour levels are at 1.64, 2.18, 3.00, 4.09, 4.77, 5.46, 6.14, 6.82, 7.50, and 8.18 Jy beam$^{-1}$ km s$^{-1}$.
c) $^{13}$CO(2--1) integrated intensity map at [61, 70] km s$^{-1}$. 
The contour levels are at (0.15, 0.2, 0.3, 0.4, 0.5, 0.6, 0.7, 0.8, 0.9, 0.98) $\times$ 45.57 Jy beam$^{-1}$ km s$^{-1}$ 
(1$\sigma$ $\sim$1.2 Jy beam$^{-1}$ km s$^{-1}$). 
d) Contour map of the C$^{18}$O(2--1) integrated intensity emission at [61.2, 69.2] km s$^{-1}$. 
The contour levels are at (0.15, 0.2, 0.3, 0.4, 0.5, 0.6, 0.7, 0.8, 0.9, 0.98) $\times$ 14.68 Jy beam$^{-1}$ km s$^{-1}$ 
(1$\sigma$ $\sim$0.45 Jy beam$^{-1}$ km s$^{-1}$). 
e) The $^{13}$CS(5--4) integrated intensity map at [61.3, 68.3] km s$^{-1}$. 
The contour levels are at (0.3, 0.45, 0.6, 0.7, 0.8, 0.9, 0.98) $\times$ 4.97 Jy beam$^{-1}$ km s$^{-1}$ 
(1$\sigma$ $\sim$0.35 Jy beam$^{-1}$ km s$^{-1}$). 
f) Contour map of the HC$_{3}$N(24--23) integrated intensity emission at [60.6, 69.6] km s$^{-1}$. 
The contour levels are at (0.22, 0.3, 0.4, 0.5, 0.6, 0.7, 0.8, 0.9, 0.98) $\times$ 3.93 Jy beam$^{-1}$ km s$^{-1}$ 
(1$\sigma$ $\sim$0.23 Jy beam$^{-1}$ km s$^{-1}$). 
g) The SiO(5--4) integrated intensity map at [62.6, 67.1] km s$^{-1}$. 
The contour levels are at (0.55, 0.6, 0.7, 0.8, 0.9, 0.98) $\times$ 0.67 Jy beam$^{-1}$ km s$^{-1}$ 
(1$\sigma$ $\sim$0.11 Jy beam$^{-1}$ km s$^{-1}$). 
h) Contour map of the SO(5$_{5}$--4$_{4}$) integrated intensity emission at [60.3, 69.3] km s$^{-1}$. 
The contour levels are at (0.3, 0.4, 0.5, 0.6, 0.7, 0.8, 0.9, 0.98) $\times$ 4.50 Jy beam$^{-1}$ km s$^{-1}$ 
(1$\sigma$ $\sim$0.33 Jy beam$^{-1}$ km s$^{-1}$). 
i) The SO(5$_{6}$--4$_{5}$) integrated intensity map at [60.6, 69.6] km s$^{-1}$. 
The contour levels are at (0.2, 0.3, 0.4, 0.5, 0.6, 0.7, 0.8, 0.9, 0.98) $\times$ 5.12 Jy beam$^{-1}$ km s$^{-1}$ 
(1$\sigma$ $\sim$0.33 Jy beam$^{-1}$ km s$^{-1}$). 
In each panel, the positions of a 6.7 GHz MME (diamond symbol) and peak positions 
of mm continuum sources (i.e., MM1--2; plus symbols) are marked. The ALMA 1.35 mm continuum contours (in dodger blue) are 
plotted at (0.15, 0.2, 0.3, 0.4, 0.5, 0.6, 0.7, 0.8, 0.9, 0.95) $\times$ 23.4 mJy beam$^{-1}$ in each panel (except panel ``b"). 
The synthesized beam is indicated in each panel (lower right corner).}
\label{fig4}
\end{figure*}

\begin{figure*}
\epsscale{1}
\plotone{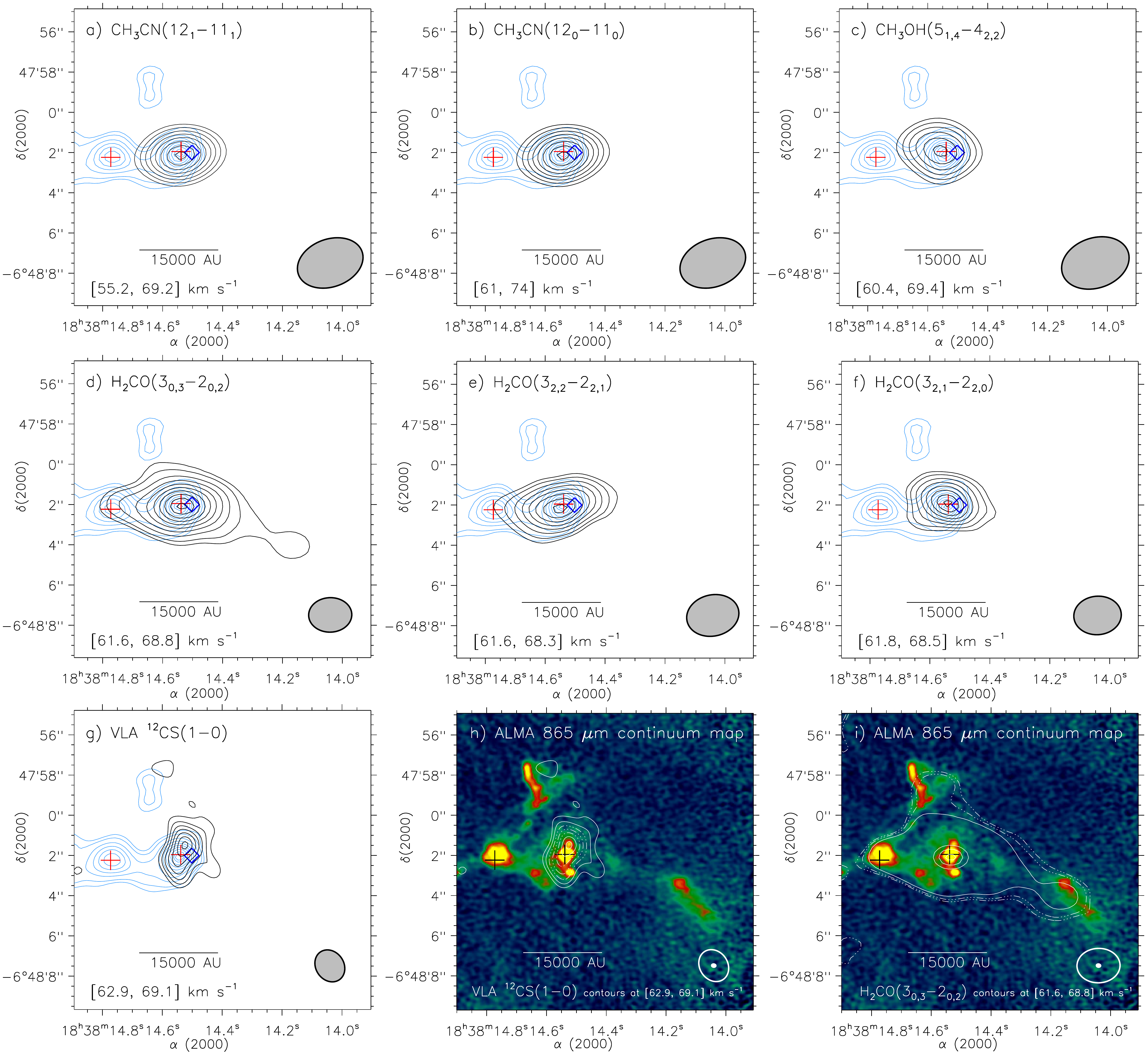}
\caption{a) Contour map of the SMA CH$_{3}$CN(12$_{1}$--11$_{1}$) integrated intensity emission. 
The contour levels are at (0.3, 0.4, 0.5, 0.6, 0.7, 0.8, 0.9, 0.98) $\times$ 5.01 Jy beam$^{-1}$ km s$^{-1}$ 
(1$\sigma$ $\sim$0.35 Jy beam$^{-1}$ km s$^{-1}$). 
b) The SMA CH$_{3}$CN(12$_{0}$--11$_{0}$) integrated intensity map. 
The contour levels are at (0.3, 0.4, 0.5, 0.6, 0.7, 0.8, 0.9, 0.98) $\times$ 4.82 Jy beam$^{-1}$ km s$^{-1}$ 
(1$\sigma$ $\sim$0.35 Jy beam$^{-1}$ km s$^{-1}$). 
c) Contour map of the SMA CH$_{3}$OH(5$_{1,4}$--4$_{2,2}$) integrated intensity emission. 
The contour levels are at (0.5, 0.6, 0.7, 0.8, 0.9, 0.98) $\times$ 1.27 Jy beam$^{-1}$ km s$^{-1}$ 
(1$\sigma$ $\sim$0.2 Jy beam$^{-1}$ km s$^{-1}$). 
d) The SMA H$_{2}$CO(3$_{0,3}$--2$_{0,2}$) integrated intensity map. 
The contour levels are at (0.14, 0.2, 0.4, 0.5, 0.6, 0.7, 0.8, 0.9, 0.98) $\times$ 3.84 Jy beam$^{-1}$ km s$^{-1}$ 
(1$\sigma$ $\sim$0.18 Jy beam$^{-1}$ km s$^{-1}$). 
e) Contour map of the SMA H$_{2}$CO(3$_{2,2}$--2$_{2,1}$) integrated intensity emission.
The contour levels are at (0.32, 0.4, 0.5, 0.6, 0.7, 0.8, 0.9, 0.98) $\times$ 1.68 Jy beam$^{-1}$ km s$^{-1}$ 
(1$\sigma$ $\sim$0.17 Jy beam$^{-1}$ km s$^{-1}$). 
f) The SMA H$_{2}$CO(3$_{2,1}$--2$_{2,0}$) integrated intensity map.
The contour levels are at (0.32, 0.4, 0.5, 0.6, 0.7, 0.8, 0.9, 0.98) $\times$ 1.72 Jy beam$^{-1}$ km s$^{-1}$ 
(1$\sigma$ $\sim$0.18 Jy beam$^{-1}$ km s$^{-1}$). 
g) Contour map of the VLA CS(1--0) integrated intensity emission. 
The contour levels are at (0.4, 0.5, 0.6, 0.7, 0.8, 0.9, 0.98) $\times$ 363 mJy beam$^{-1}$ km s$^{-1}$ 
(1$\sigma$ $\sim$47 mJy beam$^{-1}$ km s$^{-1}$).
h) Overlay of the VLA CS(1--0) emission contours on the ALMA 865 $\mu$m continuum map.
The contour levels are at (0.4, 0.5, 0.6, 0.7, 0.8, 0.9, 0.98) $\times$ 362 mJy beam$^{-1}$ km s$^{-1}$ 
(1$\sigma$ $\sim$47 mJy beam$^{-1}$ km s$^{-1}$).
i) Overlay of the H$_{2}$CO(3$_{0,3}$--2$_{0,2}$) emission contours on the ALMA 865 $\mu$m continuum map. 
The contour levels are at (0.08, 0.1, 0.14, 0.8, 0.9, 0.98) $\times$ 3.84 Jy beam$^{-1}$ km s$^{-1}$ 
(1$\sigma$ $\sim$0.18 Jy beam$^{-1}$ km s$^{-1}$). The plus and diamond symbols are the same as in Figure~\ref{fig4}. 
In panels ``a--g", the ALMA 1.35 mm continuum contours (in dodger blue) are 
plotted at (0.15, 0.2, 0.3, 0.4, 0.5, 0.6, 0.7, 0.8, 0.9, 0.95) $\times$ 23.4 mJy beam$^{-1}$. 
The synthesized beam is indicated in each panel (lower right corner).} 
\label{fig5}
\end{figure*}

In Figures~\ref{fig4} and~\ref{fig5} we present the integrated intensity maps of different molecular emission (spatial resolution $\sim$1$''$--3\rlap.{$''$}5) traced by the SMA and VLA. 
We do not detect any NH$_{3}$ or SiO(1--0) emission with the VLA. 
Only one line (i.e., CS(1--0)) from the VLA is presented in this work.

The lines detected by the SMA facility are CO(2--1), $^{13}$CO(2--1), C$^{18}$O(2--1), $^{13}$CS(5--4), HC$_{3}$N(24--23), 
SiO(5--4), SO(5$_{5}$--4$_{4}$), SO(5$_{6}$--4$_{5}$), CH$_{3}$CN(12$_{1}$--11$_{1}$), CH$_{3}$CN(12$_{0}$--11$_{0}$), 
CH$_{3}$OH(5$_{1,4}$--4$_{2,2}$), H$_{2}$CO(3$_{0,3}$--2$_{0,2}$), H$_{2}$CO(3$_{2,2}$--2$_{2,1}$), and H$_{2}$CO(3$_{2,1}$--2$_{2,0}$). 
We find compact emission coincident with MM1 in many molecular species. 
The SMA detected dense/hot gas tracers CH$_3$CN, CH$_3$OH, HC$_3$N, $^{13}$CS, shock tracer SiO, two SO transitions, and H$_2$CO. 
The detections of different sub-mm molecular lines suggest the presence of 
a hot molecular core associated with W42-MME. 
The analysis of the gas temperature of MM1 is presented in Section~\ref{frottemp}.

The maps of three CO isotopologues $^{12}$CO, $^{13}$CO, and C$^{18}$O are presented in 
Figures~\ref{fig4}a,~\ref{fig4}c, and~\ref{fig4}d, respectively. 
Figure~\ref{fig4}b shows a molecular outflow traced using the CO(2--1) line; 
the outflow is centered at the continuum peak MM1. We find a very compact morphology of the 
shock tracer SiO in the direction of continuum source MM1 
(see Figure~\ref{fig4}g), indicating the presence of shocked gas. 
However, in other outflow tracers (e.g., CO, SO, CS), the emission is slightly more extended than in SiO, 
suggesting that the ambient gas might have been entrained by 
the outflow/jet from W42-MME.

Using the VLA, we detected the CS(1--0) line, which is presented in Figure~\ref{fig5}g. 
The CS(1--0) emission contours are overlaid on the ALMA continuum map at 865 $\mu$m 
(see Figure~\ref{fig5}h), and appear to enclose the dusty envelope (see also Figure~\ref{fig3}d). 
Figure~\ref{fig5}i displays the overlay of 
the SMA H$_{2}$CO(3$_{0,3}$--2$_{0,2}$) emission contours on the ALMA continuum map at 865 $\mu$m, 
showing all six ALMA continuum sources distributed within the extended H$_{2}$CO emission.
\subsubsection{Molecular Lines from ALMA band-7}
\label{sec:linez}
\begin{figure*}
\epsscale{1}
\plotone{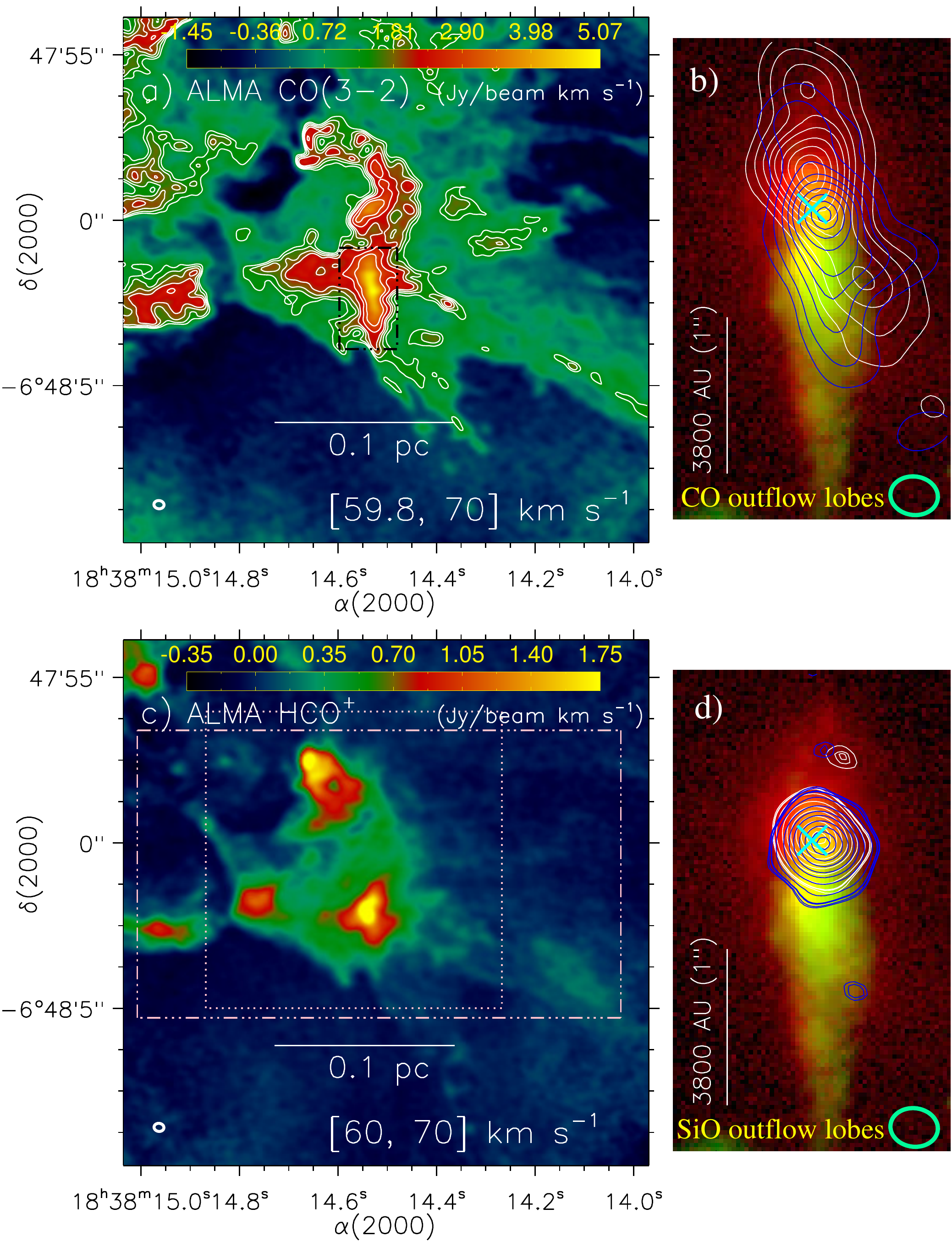}
\caption{a) Integrated intensity map of the ALMA CO(3--2) emission at [59.8, 70] km s$^{-1}$. 
The synthesized beam is 0\rlap.{$''$}31 $\times$ 0\rlap.{$''$}25, P.A. = 83$\degr$.2 (lower left corner). 
The CO(3--2) emission contours are also shown with the levels of 1.27, 1.52, 1.78, 2.03, 2.54, and 3.04 Jy beam$^{-1}$ km s$^{-1}$. 
The dotted-dashed box (in black) encompasses the area shown in Figures~\ref{fig6}b and~\ref{fig6}d.  
b) A zoomed-in view of a two-color composite NACO image around MM1a (see also Figure~\ref{fig3}d). 
The panel displays the outflow lobes of the CO(3--2) emission (redshifted component at [75, 104] km s$^{-1}$; blueshifted component at [30, 54] km s$^{-1}$). 
The contours are at (0.04, 0.1, 0.2, 0.3, 0.4, 0.55, 0.7, 0.85, 0.95) $\times$ peak value (i.e., 
2.226 Jy beam$^{-1}$ km s$^{-1}$ for redshift component and 2.846 Jy beam$^{-1}$ km s$^{-1}$ for blueshift component). 
c) Integrated intensity map of the ALMA HCO$^{+}$ emission at [60, 70] km s$^{-1}$.
The synthesized beam is 0\rlap.{$''$}30 $\times$ 0\rlap.{$''$}24, P.A. = 82$\degr$.1 (lower left corner). 
The dotted-dashed box (in pink) encompasses the area shown in Figures~\ref{fig7}a--d.
The dotted box (in pink) encompasses the area shown in Figures~\ref{fig8}a--d. 
d) The panel shows the outflow lobes of the SiO(8--7) emission (redshifted component at [73, 90] km s$^{-1}$; blueshifted component at [40, 55] km s$^{-1}$). 
The contours of redshifted component are at (0.05, 0.06, 0.063, 0.1, 0.2, 0.3, 0.4, 0.55, 0.7, 0.85, 0.95) 
$\times$ 381 mJy beam$^{-1}$ km s$^{-1}$. 
The contours of blueshifted component are at (0.045, 0.05, 0.06, 0.063, 0.1, 0.2, 0.3, 0.4, 0.55, 0.7, 0.85, 0.95) $\times$ 389 mJy beam$^{-1}$ km s$^{-1}$. 
In panels ``b" and ``d", a multiplication symbol indicates the location of the continuum source ``A".} 
\label{fig6}
\end{figure*}

\begin{figure*}
\epsscale{1.1}
\plotone{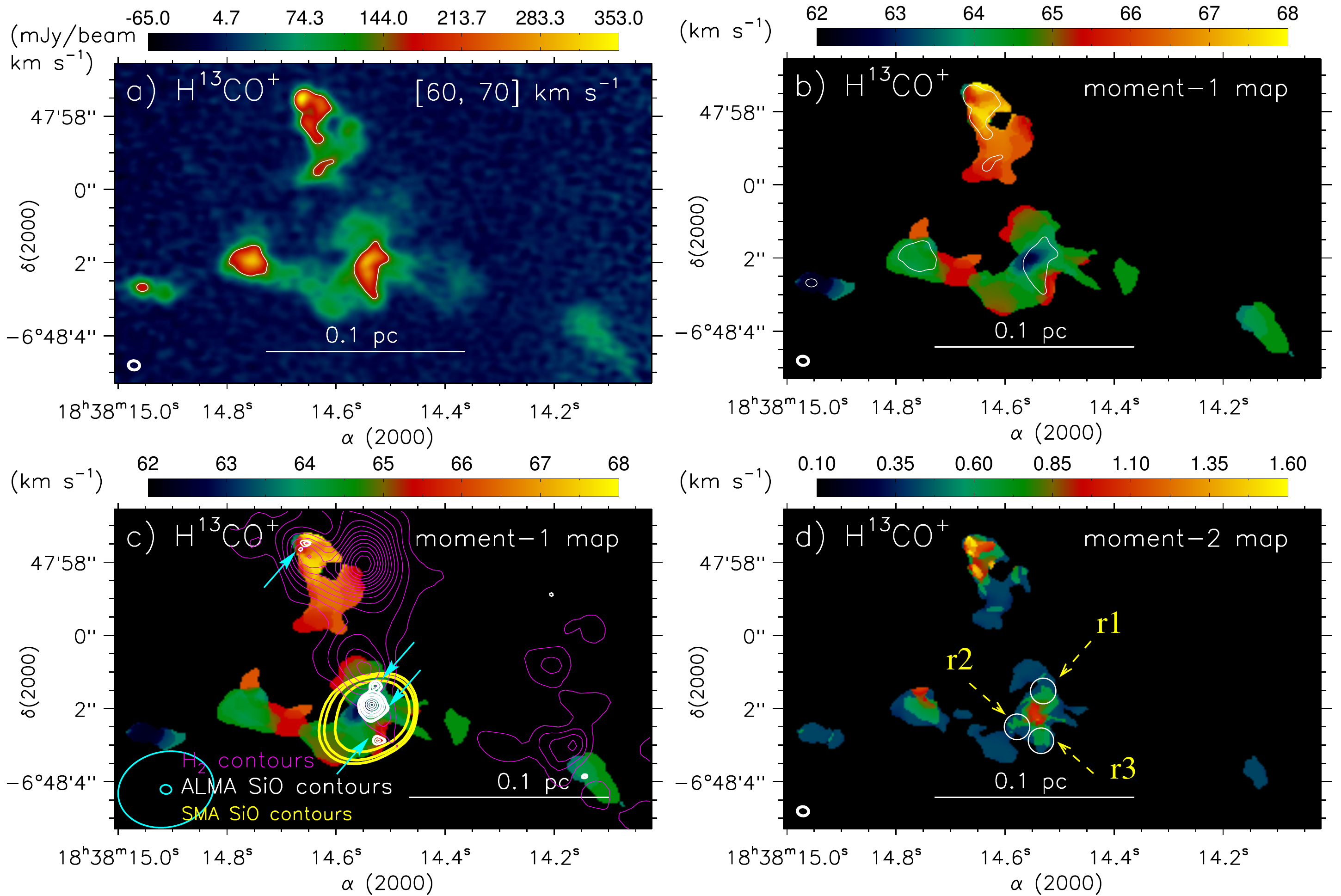}
\caption{a) Integrated intensity map of the ALMA H$^{13}$CO$^{+}$ emission at [60, 70] km s$^{-1}$ 
The synthesized beam is 0\rlap.{$''$}31 $\times$ 0\rlap.{$''$}25, P.A. = 83$\degr$.2 (lower left corner). 
b) H$^{13}$CO$^{+}$ moment-1 map. 
c) Overlay of the SiO and H$_{2}$ emission contours on the H$^{13}$CO$^{+}$ moment-1 map. 
The SMA SiO emission is shown by yellow thick contours (see Figure~\ref{fig4}g), while magenta thin contours are H$_{2}$ emission (see Figure~\ref{fig1}a). 
The contours of the ALMA SiO(8--7) emission (in white) are at 24.16, 27.61, 30.20, 38.83, 
43.14, 47.45, 86.28, 129.42, 172.56, 258.84, 345.12, 517.67, 690.23, 819.65 mJy beam$^{-1}$ km s$^{-1}$ (1$\sigma$ $\sim$4.2 mJy beam$^{-1}$ km s$^{-1}$). 
d) H$^{13}$CO$^{+}$ moment-2 map. Three small regions (r1, r2, and r3) are indicated by circles (radius $\sim$0\rlap.{$''$}35), where the profiles of the H$^{13}$CO$^{+}$ emission are studied (see Section~\ref{llsub:inner1}). 
In panels ``a" and ``b", a solid contour (in white) of the H$^{13}$CO$^{+}$ emission is also shown with a level of 150 mJy beam$^{-1}$.} 
\label{fig7}
\end{figure*}

In this section we examine CO(3--2), HCO$^{+}$(4--3), H$^{13}$CO$^{+}$(4--3), and SiO(8--7) lines from the ALMA band-7. 
In Figures~\ref{fig6}a,~\ref{fig6}c, and~\ref{fig7}a, we display the integrated intensity maps of the CO(3--2), HCO$^{+}$, and 
H$^{13}$CO$^{+}$ emission (resolution $\sim$0\rlap.{$''$}3) of an area around W42-MME, respectively. 
A comparison of the morphology of the integrated line emission with the dust 
shows a similar appearance, suggesting that these line data can be utilized to study motions of the circumstellar 
materials around W42-MME. Such a study cannot be done using the SMA, ALMA band-6, or VLA data, 
which have relatively lower resolution (i.e., $\sim$1$''$--3\rlap.{$''$}5; C.F. Table~\ref{tab1}). 
The H$^{13}$CO$^{+}$ line traces denser regions compared to the CO(3--2) and HCO$^{+}$ lines.  

Figures~\ref{fig6}b and~\ref{fig6}d show the overlay of the CO(3--2) and SiO(8--7) outflow lobes on a two-color composite 
NACO image (L$'$(red) + K$_{s}$ (green)) around MM1a, respectively. 
The CO outflow lobes are studied in velocity ranges of [30, 54] and [75, 104] km s$^{-1}$, while the SiO outflow lobes 
are shown in velocity ranges of [40, 55] and [73, 90] km s$^{-1}$. 
Both the molecular outflows are centered at the continuum source ``A" (see a multiplication symbol in Figures~\ref{fig6}b and~\ref{fig6}d). 
The SiO outflow lobes are spatially concentrated toward source ``A".
The SiO outflow lobes are more compact (extent $\sim$3500 AU) than the CO outflow lobes (extent $\sim$9000 AU). 
It seems that the CO outflow is in the plane of the sky, while the SiO outflow is to the plane of the sky. 
Furthermore, the spatial extent of the CO outflow lobes also hints the outflow cavity walls, which might also be outlined by the VLA CS emission. 

In the maps of the HCO$^{+}$ and H$^{13}$CO$^{+}$ emission, the molecular gas toward the continuum sources (MM1a, MM1b, MM2, MM3, MM4, and MM5) is seen. The highest intensity is found toward the continuum source MM3 which shows a bow-like appearance. 
Figure~\ref{fig7}b presents the moment-1 map of the H$^{13}$CO$^{+}$ emission, showing 
the intensity-weighted mean velocity of the emitting gas. The gas associated with the continuum source MM3 appears redshifted 
with respect to other continuum sources. 
A noticeable velocity difference can also be seen toward the continuum source MM1a (see the H$^{13}$CO$^{+}$ emission contour in Figure~\ref{fig7}b). In Figure~\ref{fig7}c, we present the overlay of the contours of the SMA SiO(5--4), ALMA SiO(8--7), and H$_{2}$ emission on the H$^{13}$CO$^{+}$ moment-1 map. 
The emission peak of the SMA SiO(5--4) is detected toward the continuum source MM1a.
However, the ALMA SiO(8--7) emission at [60.8, 71] km s$^{-1}$ is seen toward the continuum sources ``A--C" and the H$_{2}$ knot (see arrows in Figure~\ref{fig7}c). 
The dusty envelope as well as continuum peaks appear to be influenced (i.e., heated) by shocks. 
Therefore, using the continuum map, mass estimations of the continuum sources distributed within the source MM1a 
may not be accurate. Furthermore, the SiO(8--7) emission is also detected in the direction of the tip of the bow-like appearance of the continuum source MM3, where the H$_{2}$ knot is seen. Figure~\ref{fig7}d displays the moment-2 map or the intensity-weighted dispersion map of the H$^{13}$CO$^{+}$ emission. In the direction of the continuum sources MM1, MM2, and MM3, we find a velocity dispersion larger than 1~km\,s$^{-1}$. 
The velocity dispersion toward the source ``A" may be related to rotation (see also Section~\ref{sub:inner2}).   

\begin{figure*}
\epsscale{1.1}
\plotone{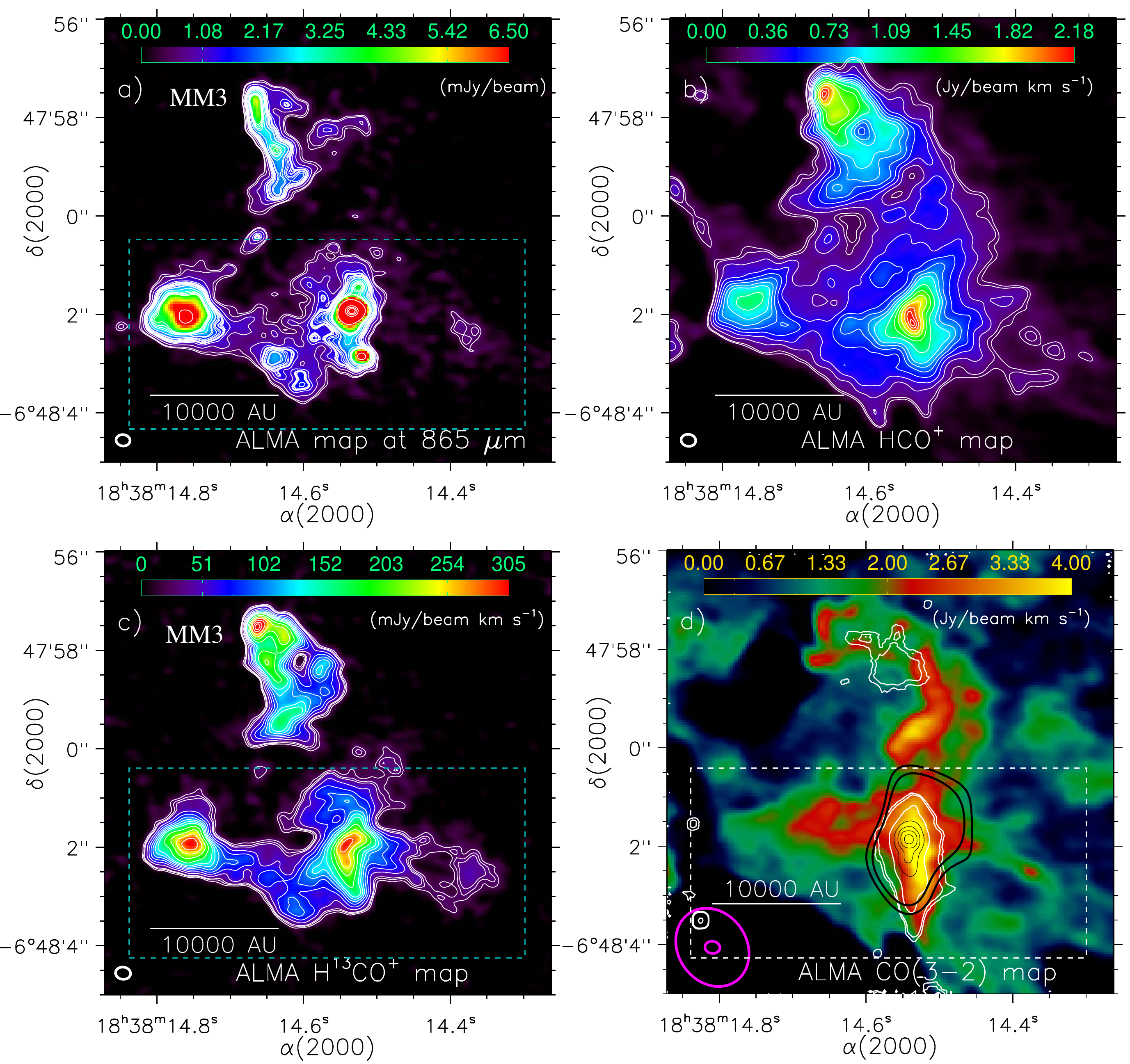}
\caption{A zoomed-in view of an area containing the mm continuum sources MM1--3 using the ALMA maps (see a dotted box in Figure~\ref{fig7}a). 
a) The panel shows the ALMA continuum map and contours at 865 $\mu$m. 
The contour levels are the same as in Figure~\ref{fig3}c. 
b) The panel displays the integrated intensity map and contours of the HCO$^{+}$ 
emission at [60, 70] km s$^{-1}$. 
The contour levels are at 
(0.085, 0.1, 0.15, 0.2, 0.25, 0.3, 0.35, 0.4, 0.5, 0.6, 0.7, 0.8, 0.9, 0.96, 0.99) $\times$ 2.19 
Jy beam$^{-1}$ km s$^{-1}$ (1$\sigma$ $\sim$0.025 Jy beam$^{-1}$ km s$^{-1}$). 
c) The panel displays the integrated intensity map and contours of the 
H$^{13}$CO$^{+}$ emission at [60, 70] km s$^{-1}$. 
The contour levels are at (0.065, 0.085, 0.1, 0.15, 0.2, 0.25, 0.3, 0.35, 0.4, 0.5, 0.6, 0.7, 0.8, 0.9, 0.96) $\times$ 
353.2 mJy beam$^{-1}$ km s$^{-1}$ (1$\sigma$ $\sim$7.5 mJy beam$^{-1}$ km s$^{-1}$). 
d) Overlay of the NACO L$'$ emission contours (see thin contours in white and black) on the integrated intensity map of the 
ALMA CO(3--2) emission at [59.8, 70] km s$^{-1}$ (see Figures~\ref{fig2}c and~\ref{fig6}a). The CO map is also overlaid with the VLA CS(1--0) emission, which is shown by thick contours (in black) at (0.52, 0.6) $\times$ 362 mJy beam$^{-1}$ km s$^{-1}$ (1$\sigma$ $\sim$47 mJy beam$^{-1}$ km s$^{-1}$). In panels a, c, and d, the dashed box highlights the elongated feature containing the continuum sources MM1a, MM1b, and MM2.} 
\label{fig8}
\end{figure*}

In Figure~\ref{fig8}a, we show the ALMA continuum map and contours at 865 $\mu$m toward 
an area containing the mm continuum sources MM1--3. 
Figures~\ref{fig8}b and~\ref{fig8}c display the integrated intensity maps and contours of the 
HCO$^{+}$ and H$^{13}$CO$^{+}$ emission at [60, 70] km s$^{-1}$, respectively. 
In Figure~\ref{fig8}d, we present the integrated intensity map of the CO(3--2) emission at [59.8, 70] km s$^{-1}$ and 
the NACO L$'$ emission contours. 
The VLA CS(1--0) emission contours (in black) are also shown in Figure~\ref{fig8}d. 
We find strong intensity of the molecular emission toward the continuum source ``A" and the dusty envelope. 
The spatial morphology of the continuum source ``A" and the dusty envelope look little different in the VLT/NACO L$'$ image, and the emission maps of CO(3--2), H$^{13}$CO$^{+}$, and HCO$^{+}$ (see Figure~\ref{fig8} and also Section~\ref{sub:inner1}).

The 865 $\mu$m continuum map and the H$^{13}$CO$^{+}$ emission map exhibit a similar morphology, 
where an emission feature (extent $\sim$0.1 pc) hosting the continuum sources MM1a, MM1b, and MM2 is evident 
(see the dotted box in Figures~\ref{fig8}a and~\ref{fig8}c). 
As mentioned earlier, the continuum source MM3 is very prominent in the continuum map and the integrated line maps 
of the H$^{13}$CO$^{+}$ and HCO$^{+}$ emission, and is associated with 
the shock-excited molecular line emission resulting from the W42-MME jet/outflow activity (see Figure~\ref{fig7}c). 
Note that the location of MM3 is far away from the extent of the ALMA CO outflow lobes. 
More discussion on these findings is given in Section~\ref{sec:disc}.  
\subsection{Gas temperature and non-thermal dispersion}
\label{frottemp}
\subsubsection{Gas temperature}
\label{frottemp1a}
In this section, the lines methyl cyanide/acetonitrile/cyanomethane, CH$_{3}$CN from ALMA band-6 and propyne/methyl acetylene, CH$_{3}$CCH from ALMA band-7 are explored for obtaining the gas temperature of the core MM1 or the continuum source ``A". 

\begin{figure*}
\epsscale{1}
\plotone{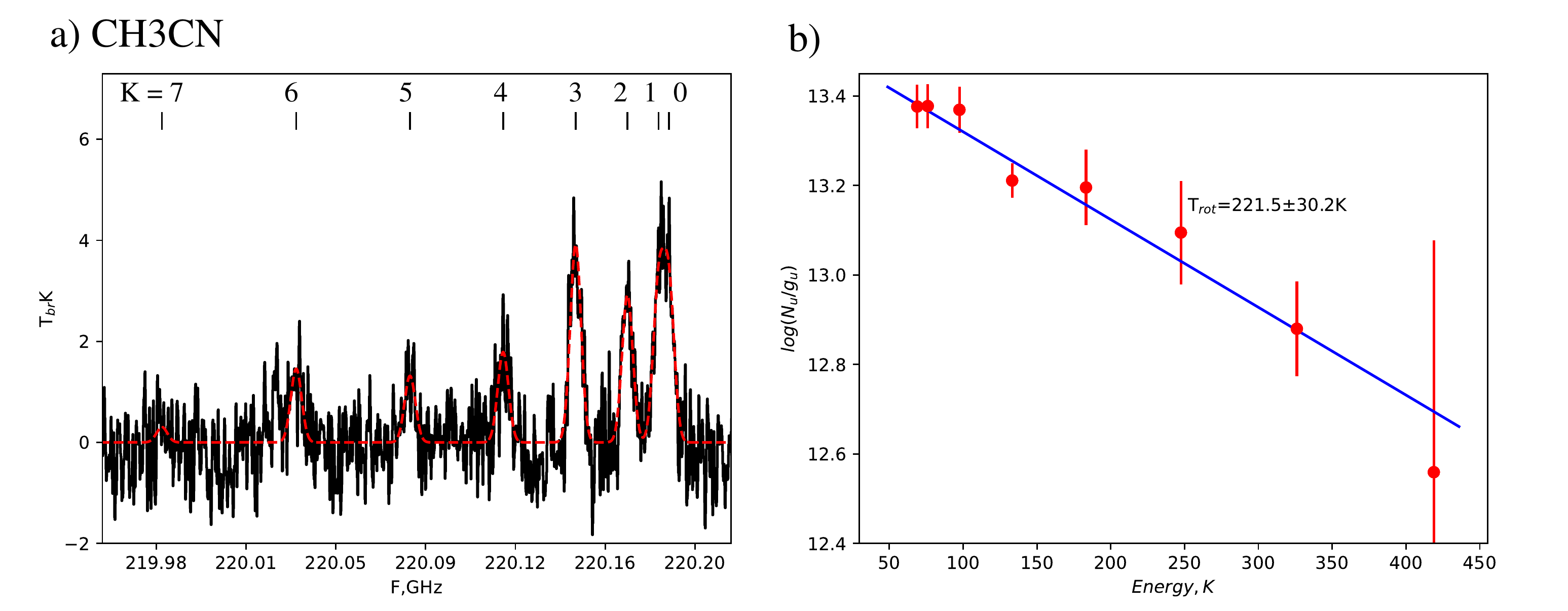}
\epsscale{1}
\plotone{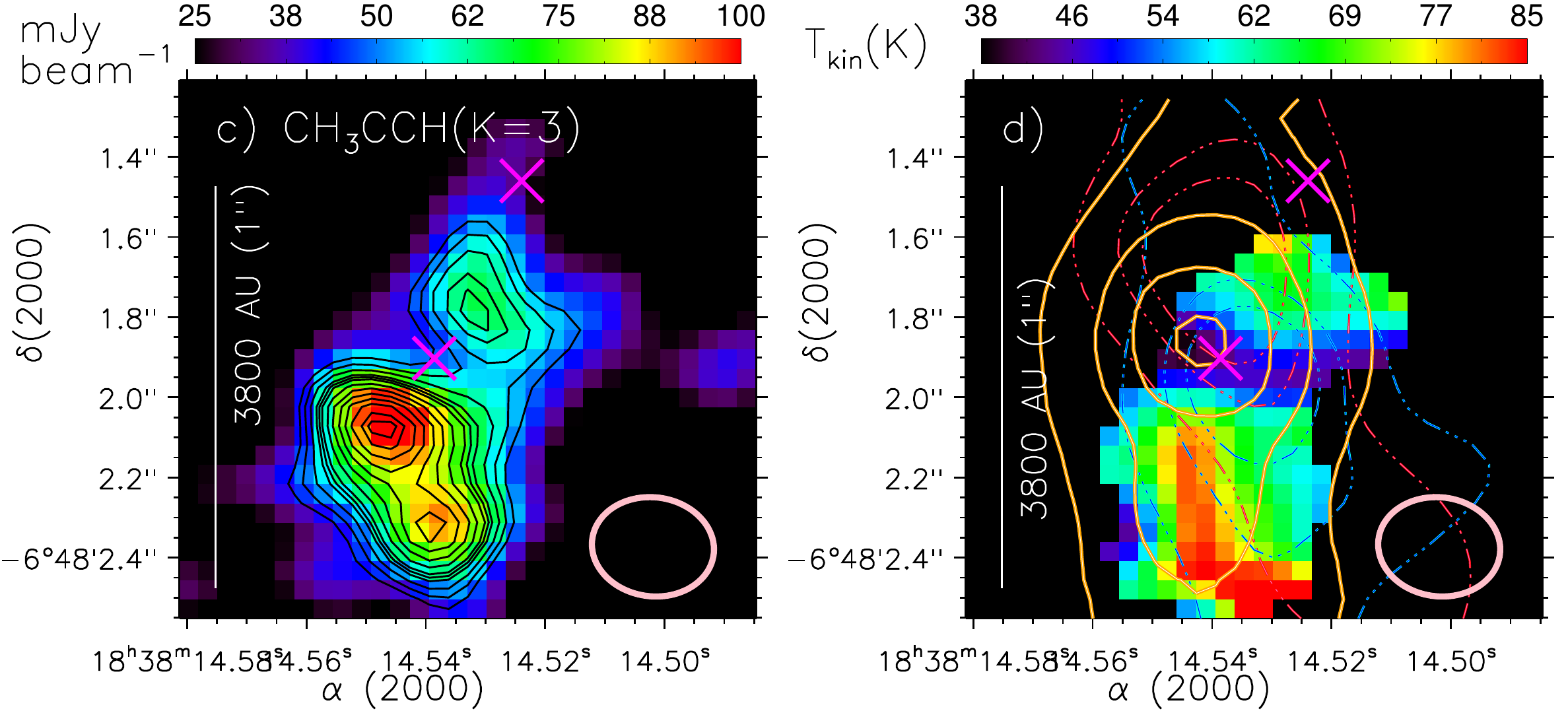}
\epsscale{0.7}
\plotone{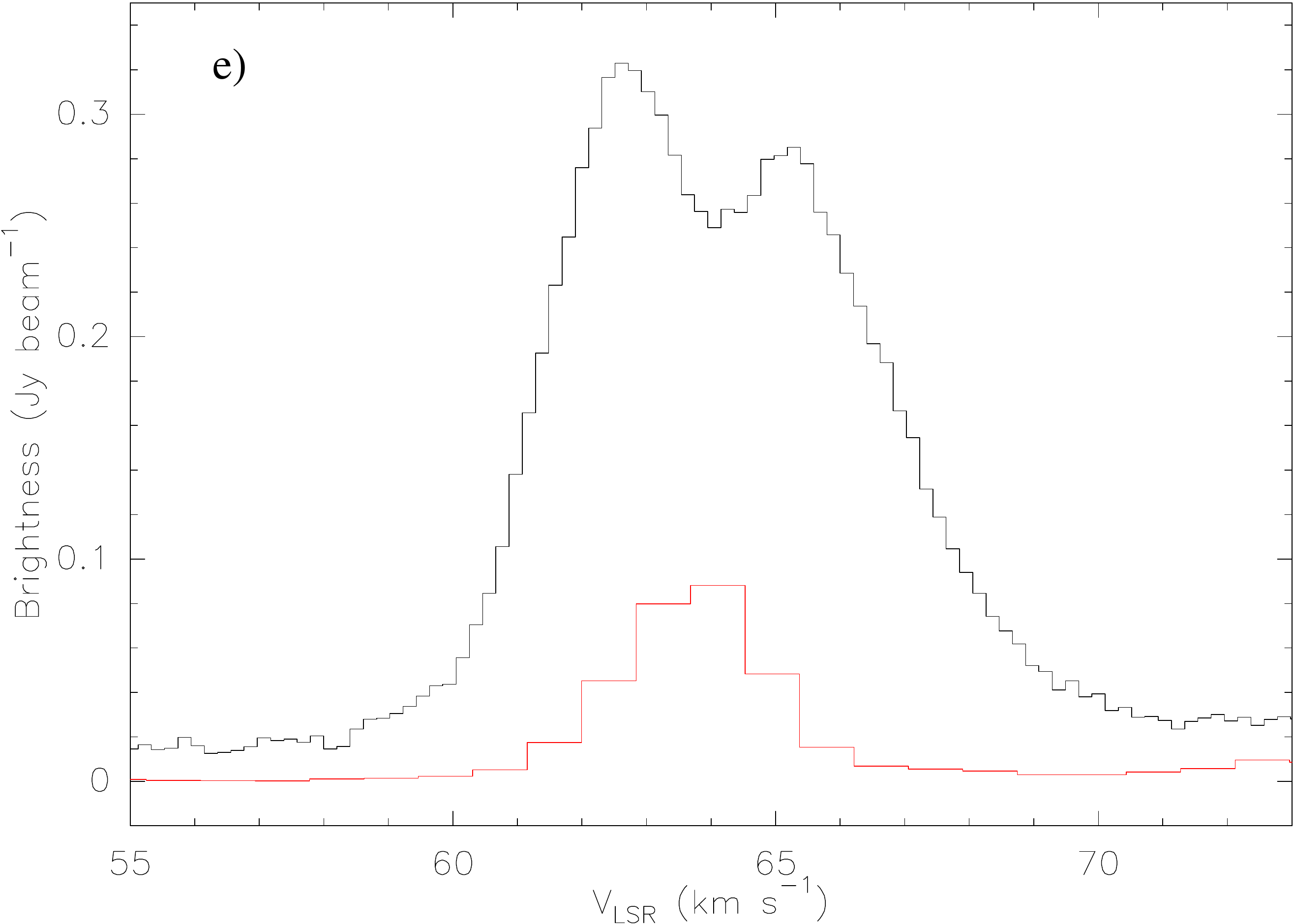}
\caption{a) The panel shows the spectra and best-fit model of the CH$_{3}$CN(12$_{K}$--11$_{K}$) $K$ =0--7 lines (resolution $\sim$1$''$) traced in the ALMA band-6. 
The observed ALMA CH$_{3}$CN transitions are plotted by the black line, 
while the synthetic spectra obtained from the best-fit model are shown by the red line. 
The observed spectra are smoothed with a hanning filter with an FWHM of $\sim$1 km s$^{-1}$.
b) The CH$_{3}$CN rotation diagram of the continuum source MM1. 
The solid line is the fit to the eight CH$_{3}$CN transitions as shown in Figure~\ref{fig12}a.
c) Intensity map and contours of the methyl acetylene (propyne) CH$_{3}$CCH ($K$ =3 transition) emission (resolution $\sim$0\rlap.{$''$}3) integrated from 
V$_\mathrm{lsr}$ = 57 to 71 km s$^{-1}$ in the direction of the continuum source MM1a. The contours (in black) of the CH$_{3}$CCH emission are at (0.45, 0.5, 0.55, 0.6, 0.625, 0.65, 0.7, 0.75, 0.8, 0.85, 0.9, 0.95, 0.98) $\times$ 105 mJy beam$^{-1}$ km s$^{-1}$ (1$\sigma$ $\sim$9 mJy beam$^{-1}$ km s$^{-1}$). 
d) The kinetic temperatures derived from the transitions of the CH$_{3}$CCH line toward the continuum source MM1a. 
The VLT/NACO L$'$ emission contours (in khaki; see also Figure~\ref{fig3}b) are also overlaid on the kinetic temperature map. 
The ALMA CO outflow lobes are also shown by dotted-dashed curves (see Figure~\ref{fig6}b). 
In panels ``c" and ``d", multiplication symbols indicate the locations of the continuum sources ``A" and ``C". 
e) Histogram profiles represent the HCO$^{+}$ emission (in black) and H$^{13}$CO$^{+}$ emission (in red).} 
\label{fig12}
\end{figure*}

We examined several transitions of the CH$_{3}$CN emission observed in ALMA band-6 (resolution $\sim$1$''$). 
Eight components of the CH$_{3}$CN $K$-ladder with $K$=0-7 are observed, and their spectra are presented in Figure~\ref{fig12}a. 
These transitions have been used to determine the rotational temperature at the location of the continuum peak MM1. 
In Figure~\ref{fig12}a, all CH$_{3}$CN transitions are simultaneously fitted by a model given in \citet{araya05}. 
The model assumes local thermodynamic equilibrium (LTE) conditions. 
In Figure~\ref{fig12}b, we present the CH$_{3}$CN rotation diagram of the continuum source MM1. The best fit yields a rotational temperature, $T_{\rm rot}$, of 221.5$\pm$30.2 K. The uncertainty in the CH$_{3}$CN temperature is 3$\sigma$. 
Using the different transitions of the SMA CH$_{3}$CN line (resolution $\sim$3$''$), we derive a 
rotational temperature of $\sim$152~K.  

Five components of the CH$_{3}$CCH emission (beam size $\sim$0\rlap.{$''$}3 $\times$ 0\rlap.{$''$}24) are detected in the ALMA band-7. These data have a better spatial resolution compared to the SMA/ALMA CH$_{3}$CN data. In Figure~\ref{fig12}c, we present the integrated intensity map and contours of the CH$_{3}$CCH ($K$ =3 transition) emission around the continuum source ``A". 
The kinetic temperature of the CH$_{3}$CCH gas is determined by the method 
of population diagrams \citep[see][for more details]{malafeev05}. 
Figure~\ref{fig12}d displays the kinetic temperature map around the source ``A". 
The temperature map is also overlaid with the ALMA CO outflow direction (i.e., NE--SW) and the NACO L$'$ emission contours. 
The locations of the continuum sources ``A" and ``C" are marked by multiplication symbols in the map. 
The range of the gas temperature is determined to be [38, 85]~K. 
A noticeable gas temperature gradient is evident toward the source ``A" or the areas 
covered by the NACO emission contours, which highlight the proposed ionized jet-like feature and the location of the MYSO. Higher gas temperatures are found toward regions located in the North and SE directions. 
The proposed ionized jet-like feature is detected with higher gas temperatures (i.e., 60--85~K; mean value $\sim$70~K). 
All these exercises suggest that the core MM1 or MM1a is heated by the MYSO W42-MME. 
One can also notice a difference in temperature estimates from the CH$_{3}$CN and CH$_{3}$CCH emission. 
It can be explained with the fact that the CH$_{3}$CN emission is produced from a warmer and inner region of the envelope than the 
CH$_{3}$CCH emission \citep[e.g.,][]{andron18}.

Taking into account the detections of several sub-mm lines and higher gas temperature, our results confirm the presence of a hot molecular core associated with the MYSO W42-MME. 
\subsubsection{Non-thermal dispersion and signature of infall motion}
\label{llsub:inner1}
Using the optically thin H$^{13}$CO$^{+}$ line, we examined the spectra toward three small regions (i.e., r1, r2, and r3) around the continuum source ``A" (see circles in Figure~\ref{fig7}d), and computed the FWHM linewidth of each observed H$^{13}$CO$^{+}$ profile (not shown here). Using the observed FWHM value, we determined the sound speed ($a_\mathrm{s}$), thermal velocity dispersion ($\sigma_{T}$), non-thermal velocity dispersion ($\sigma_\mathrm{NT}$), Mach number ($M$ = $\sigma_\mathrm{NT}$/$a_\mathrm{s}$), and ratio of thermal to non-thermal gas pressure \citep[$R_\mathrm{p}$ = ${a^2_\mathrm{s}}/{\sigma^2_\mathrm{NT}}$; see][for more details]{lada03}. The sound speed ($a_\mathrm{s}$ = $(k T_\mathrm{kin}/\mu m_\mathrm{H})^{1/2}$) is estimated for $\mu$ = 2.37 (approximately 70\% H and 28\% He by mass) and a range of temperature (i.e., $T_\mathrm{kin}$ = [40, 70] K).
The non-thermal velocity dispersion is defined as:
\begin{equation}
\sigma_\mathrm{NT} = \sqrt{\frac{\Delta V^2}{8\ln 2}-\frac{k T_\mathrm{kin}}{30 m_\mathrm{H}}} = \sqrt{\frac{\Delta V^2}{8\ln 2}-\sigma_\mathrm{T}^{2}} ,
\label{sigmanonthermal}
\end{equation}
where $\Delta V$ is the measured linewidth of the observed H$^{13}$CO$^{+}$ profile, 
$\sigma_\mathrm{T}$ (= $(k T_\mathrm{kin}/30 m_\mathrm{H})^{1/2}$) is the thermal broadening for H$^{13}$CO$^{+}$. 
In the direction of regions r1, r2, and r3, the value of non-thermal velocity dispersion 
is determined to be 0.92(0.91), 0.93(0.92), and 1.0(1.0) km s$^{-1}$ at $T_\mathrm{kin}$ = 40(70) K, respectively. 
Using the value of $T_\mathrm{kin}$ = 40(70) K, we obtain the sound speed to be 0.37(0.49) km s$^{-1}$ toward the regions r1, r2, and r3. 
In the direction of the regions r1, r2, and r3, the Mach number is estimated to be 2.5(1.9), 2.5(1.9), and 2.8(2.1) 
at $T_\mathrm{kin}$ = 40(70) K, respectively. 
For the value of $T_\mathrm{kin}$ = 40(70) K, the $R_\mathrm{p}$ is estimated to be 0.16(0.29), 0.16(0.28), and 0.12(0.22) 
toward the regions r1, r2, and r3, respectively. 
Based on these derived physical parameters, we suggest that non-thermal pressure and supersonic non-thermal motions (e.g., turbulence, outflows, shocks, and/or magnetic fields) are dominant in these regions, which are not exactly coincident with the continuum source ``A" (see circles in Figure~\ref{fig7}d). 

In Figure~\ref{fig12}e, we present the profiles of the H$^{13}$CO$^{+}$(4--3) emission (in red) and HCO$^{+}$(4--3) emission (in black) toward the continuum source ``A". 
A single peak is seen in the optically thin H$^{13}$CO$^{+}$ line, while a red-shifted self-absorption dip is detected 
in the optically thick HCO$^{+}$ line at the W42-MME position. These profiles may indicate the signatures of infall toward the continuum source ``A", although this is not certain since the HCO$^{+}$ profile varies significantly across the source, which can be caused by a complicated morphology of the HCO$^{+}$ distribution. 
\subsection{Multiline spectral imaging view of continuum source MM1a}
\label{sub:inner1}
\begin{figure*}
\epsscale{0.7}
\plotone{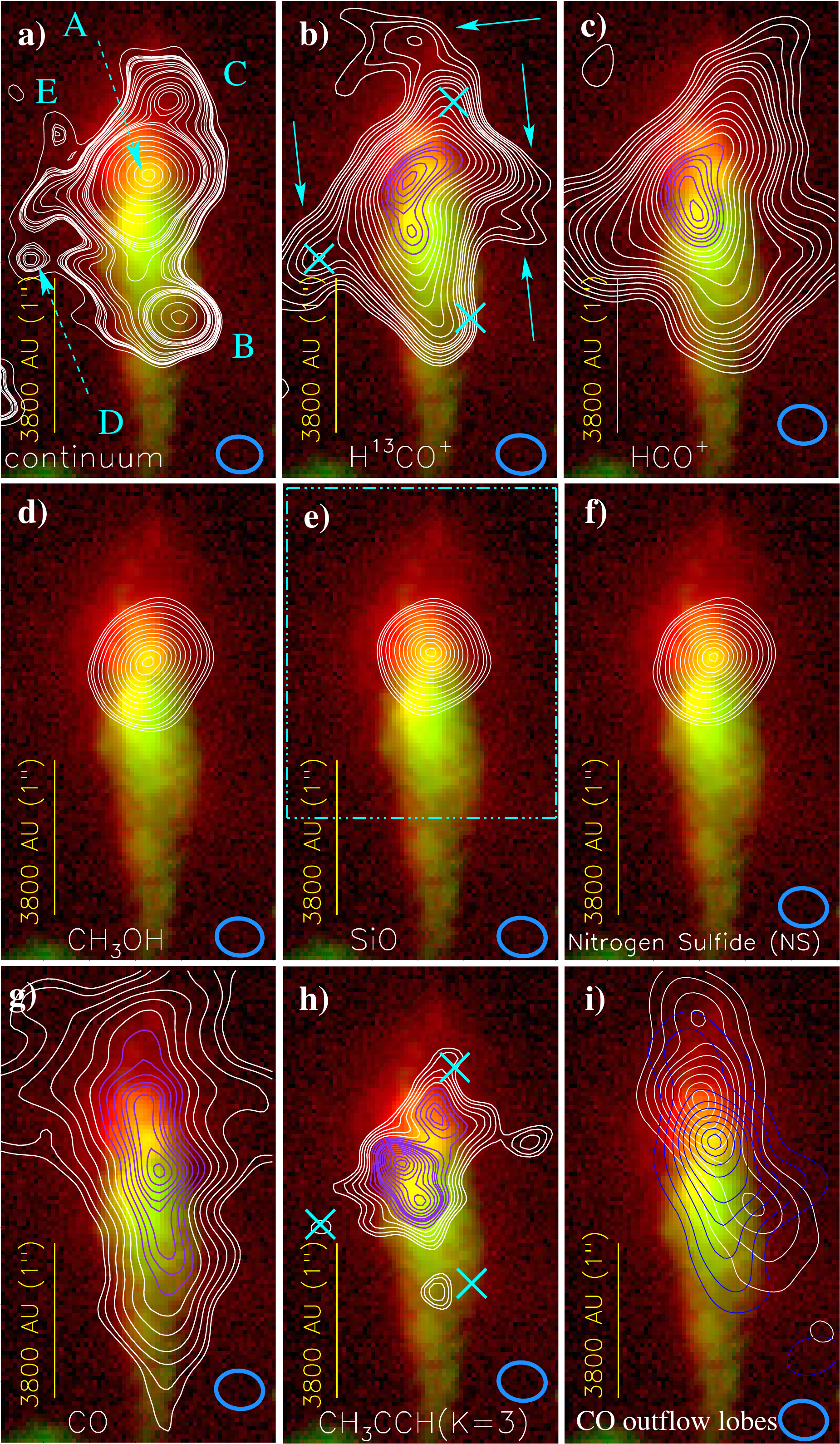}
\caption{A zoomed-in view of a two-color composite NACO map (L$'$ (red) and K$_{s}$ (green) images) 
around MM1a (see also Figure~\ref{fig3}d). 
a) The ALMA continuum emission contours at 865 $\mu$m are presented, and are the same as in Figure~\ref{fig3}b. 
b) The panel shows the integrated intensity contours of the H$^{13}$CO$^{+}$ emission at [60, 70] km s$^{-1}$. 
The contour levels are at (0.28, 0.3, 0.32, 0.35, 0.38, 0.4, 0.42, 0.45, 0.5, 0.55, 0.6, 0.65, 0.7, 0.75, 
0.8, 0.85, 0.88, 0.92, 0.96, 0.99) $\times$ 301.6 mJy beam$^{-1}$ km s$^{-1}$. 
c) The panel displays the integrated intensity contours of the HCO$^{+}$ emission at [60, 70] km s$^{-1}$. 
The contour levels are at (0.28, 0.3, 0.32, 0.35, 0.38, 0.4, 0.42, 0.45, 0.5, 0.55, 0.6, 0.65, 0.7, 0.75, 
0.8, 0.85, 0.88, 0.92, 0.96, 0.99) $\times$ 2.18 mJy beam$^{-1}$ km s$^{-1}$.
d) The panel displays the integrated intensity contours of the CH$_{3}$OH emission at [56.6, 76.2] km s$^{-1}$. 
The contours of the CH$_{3}$OH emission are at (0.06, 0.08, 0.1, 0.15, 0.2, 0.3, 0.4, 0.5, 0.6, 0.7, 0.8, 0.9, 0.98) $\times$ 2.02 Jy beam$^{-1}$ km s$^{-1}$ (1$\sigma$ $\sim$13.7 mJy beam$^{-1}$ km s$^{-1}$).  
e) The panel shows the integrated intensity contours of the SiO emission at [60.8, 71] km s$^{-1}$. 
The contours of the SiO emission are at (0.06, 0.08, 0.1, 0.15, 0.2, 0.3, 0.4, 0.5, 0.6, 0.7, 0.8, 0.9, 0.98) 
$\times$ 860.7 mJy beam$^{-1}$ km s$^{-1}$ (1$\sigma$ $\sim$4.2 mJy beam$^{-1}$ km s$^{-1}$). 
The dotted-dashed box (in cyan) encompasses the area shown in all panels of Figure~\ref{fig10}. 
f) The panel displays the integrated intensity contours of the Nitrogen Sulfide (NS) emission at [55, 70.2] km s$^{-1}$. 
The contours of the NS emission are at (0.06, 0.08, 0.1, 0.15, 0.2, 0.3, 0.4, 0.5, 0.6, 0.7, 0.8, 0.9, 0.98) $\times$ 2.06 Jy beam$^{-1}$ km s$^{-1}$ (1$\sigma$ $\sim$7.7 mJy beam$^{-1}$ km s$^{-1}$).  
g) The panel displays the integrated intensity contours of the CO emission at [59.8, 70] km s$^{-1}$. 
The contours of the CO emission are at (0.42, 0.45, 0.5, 0.55, 0.6, 0.65, 0.7, 0.75, 0.8, 0.85, 0.88, 0.92, 0.96, 0.99) 
$\times$ 5.06 Jy beam$^{-1}$ km s$^{-1}$.
h) The panel presents the integrated intensity contours of the CH$_{3}$CCH ($K$ =3 transition) emission at [57, 71] km s$^{-1}$ (see also Figure~\ref{fig12}c). The contours of the CH$_{3}$CCH emission are at (0.26, 0.3, 0.35, 0.4, 0.45, 0.5, 0.55, 0.6, 0.625, 0.65, 0.7, 0.75, 0.8, 0.85, 0.9, 0.95, 0.98) $\times$ 105 mJy beam$^{-1}$ km s$^{-1}$ (1$\sigma$ $\sim$9 mJy beam$^{-1}$ km s$^{-1}$). 
i) The panel displays the outflow lobes of the CO emission, which are the same as in Figure~\ref{fig6}b. 
In panels ``b" and ``h", multiplication symbols (in cyan) show the locations of the continuum sources ``B--D".}
\label{fig9}
\end{figure*}

In the direction of the continuum source MM1a, in Figures~\ref{fig9}a--\ref{fig9}i, 
we present a zoomed-in view of a two-color composite NACO map (L$'$ (red) and K$_{s}$ (green) images) 
overlaid with the emission contours of the 865 $\mu$m continuum, H$^{13}$CO$^{+}$(4--3), HCO$^{+}$(4--3), CH$_{3}$OH, SiO(8--7), 
nitrogen sulfide (NS), CO(3--2), CH$_{3}$CCH (k=3), and CO outflow lobes, respectively. 
Some of these lines (i.e., H$^{13}$CO$^{+}$, CH$_{3}$OH, and NS) are known as dense gas tracers. 
Figures~\ref{fig9}a,~\ref{fig9}e, and~\ref{fig9}i are the same as presented in Figures~\ref{fig3}d,~\ref{fig7}c, and~\ref{fig6}b, respectively, which are shown here only for a comparison purpose. 

In Figure~\ref{fig9}a, the continuum source ``A" is seen almost at the center of the dusty envelope, and is surrounded by four continuum peaks ``B--E". In the direction of the proposed infrared envelope/outflow cavity, the outflow cavity walls are depicted by the CO emission and the CO outflow lobes (see Figures~\ref{fig9}g and~\ref{fig9}i). 
The outer contours of the H$^{13}$CO$^{+}$ and HCO$^{+}$ emission display narrow molecular structures (see arrows in Figure~\ref{fig9}b), 
which may show the outflow cavity walls (see Figures~\ref{fig9}b and~\ref{fig9}c). 
In Figure~\ref{fig9}b, multiplication symbols (in cyan) show the locations of the continuum sources (i.e., B, C, and D), which are interestingly seen toward 
narrow structures of the H$^{13}$CO$^{+}$ emission.

The continuum source ``A" is well traced in the shock gas tracer SiO(8--7) and the dense gas tracers NS and CH$_{3}$OH. 
The spatial morphology of the continuum source ``A" appears similar in the maps of the CH$_{3}$OH, SiO(8--7), and NS emission. 
Note that the ALMA SiO data also reveal the compact SiO outflow concentrated toward the source ``A" (see Figure~\ref{fig6}d). 

The H$^{13}$CO$^{+}$, HCO$^{+}$, CO, and CH$_{3}$CCH emission contours with higher intensities are shown by magenta color, and are seen 
toward the continuum source ``A". 
The peak of the CH$_{3}$CCH emission ($K$ = 3 component) lies slightly to the south of ``A" and peaks toward 
the proposed infrared jet-like feature. 
The H$^{13}$CO$^{+}$ emission contours are distributed in the northwest to southeast direction, appearing like a flattened/elongated feature. 

\begin{figure*}
\epsscale{1}
\plotone{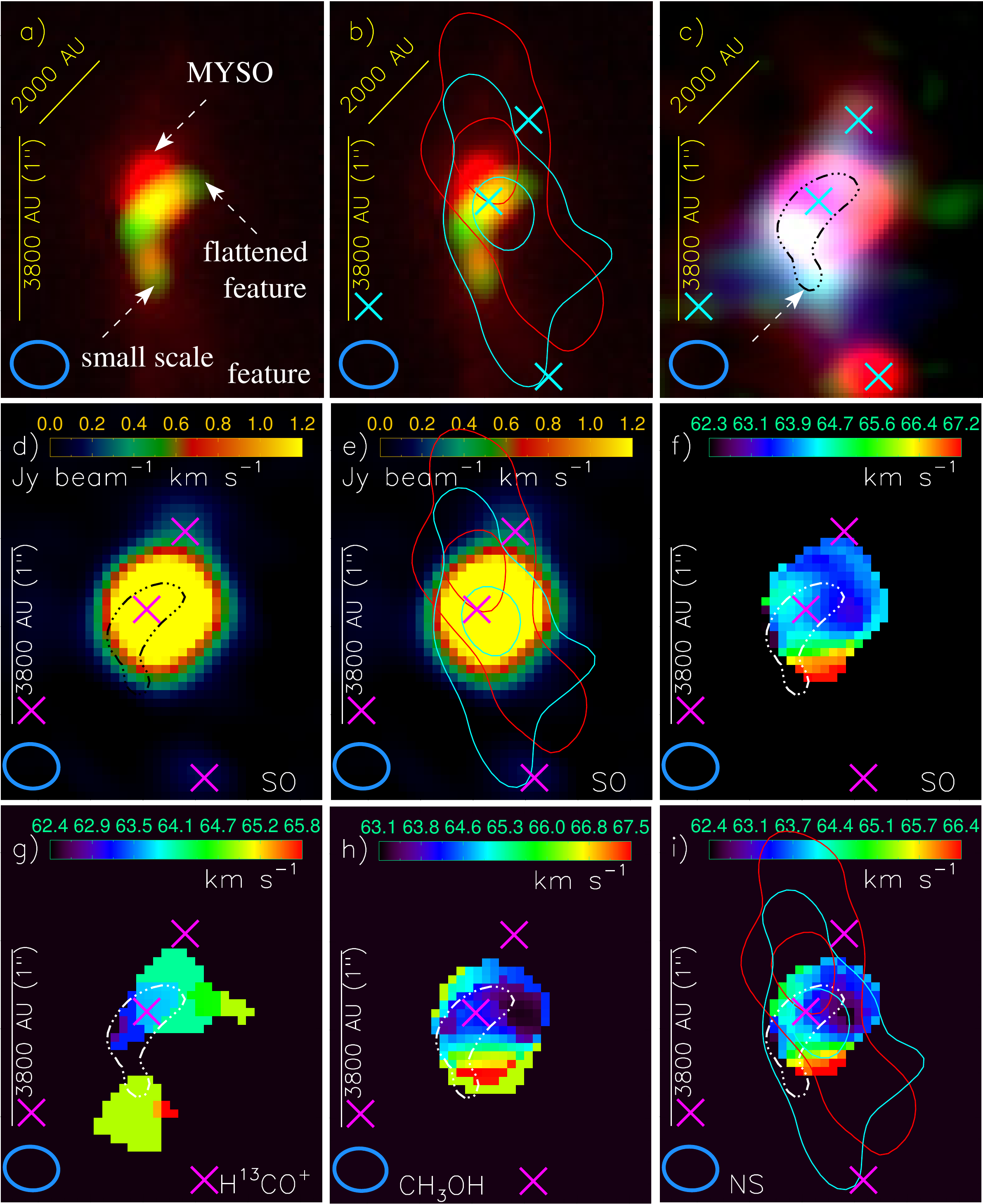}
\caption{A zoomed-in view of an area hosting the continuum sources ``A--D" (see a dotted-dashed box in Figure~\ref{fig9}e).
a) The panel shows a two color-composite image (NACO L$'$ band (red) + H$^{13}$CO$^{+}$ (green)). 
b) The panel is the same as Figure~\ref{fig10}a but overlaid with the CO outflow lobes. 
c) The panel displays a three color-composite image (ALMA continuum map at 865 $\mu$m (red) + CH$_{3}$CCH (green) + HCO$^{+}$ (blue)). 
d) The panel presents the integrated intensity map of the Sulphur monoxide (SO) 8(8)--7(7) emission at [50, 80] km s$^{-1}$. 
e) The panel is the same as Figure~\ref{fig10}d but overlaid with the CO outflow lobes. 
f) The panel shows the SO moment-1 map. 
g) The panel displays the H$^{13}$CO$^{+}$ moment-1 map. 
h) The panel presents the CH$_{3}$OH moment-1 map. 
i) Overlay of the CO outflow lobes on the Nitrogen Sulfide (NS) moment-1 map. 
In panels``b--i", multiplication symbols indicate the positions of the continuum 
sources ``A--D" (see also Figures~\ref{fig9}a and~\ref{fig9}b). 
In panels``c--d" and ``f--i", a dotted-dashed contour highlights the location of the flattened feature and 
the small scale feature as indicated in Figure~\ref{fig10}a. 
The CO outflow lobes (redshifted and blueshifted emission) are taken from Figure~\ref{fig9}h. 
Each moment-1 map is produced at the higher value of the cutoff level.}
\label{fig10}
\end{figure*}

In Figure~\ref{fig10}a, we display a two color-composite map (NACO L$'$ band (red) + H$^{13}$CO$^{+}$ (green)) toward an area containing the continuum sources ``A--D" (see a dotted-dashed box in Figure~\ref{fig9}e), strongly showing the flattened/elongated feature (extent $\sim$2000 AU) in the H$^{13}$CO$^{+}$ emission. 
The MYSO W42-MME is almost seen at the center of the flattened feature. 
Figure~\ref{fig10}b is the same as Figure~\ref{fig10}a, but the color-composite map is overlaid with the CO outflow lobes. 
The peak positions of the continuum sources (A--D) are also shown in Figure~\ref{fig10}b. 

The elongation of the H$^{13}$CO$^{+}$ emission hints that the feature is located at a large inclination. 
The orientation of the CO outflow lobes is also perpendicular to the H$^{13}$CO$^{+}$ flattened feature (see Figure~\ref{fig10}b). 
It has been suggested that outflows/jets are always launched perpendicular to the disk plane \citep{monin07}.
Hence, the flattened feature could be an accretion disk-like feature around the MYSO W42-MME. 
Figure~\ref{fig10}c displays a three color-composite image (ALMA continuum map at 865 $\mu$m (red) + CH$_{3}$CCH (green) + HCO$^{+}$ (blue)) overlaid with the H$^{13}$CO$^{+}$ emission, illustrating the association of the H$^{13}$CO$^{+}$ flattened feature with the continuum emission and HCO$^{+}$. However, the flattened/elongation morphology is not seen in the continuum map. 
In Figures~\ref{fig10}d and~\ref{fig10}e, we present the moment-0 map of the Sulphur monoxide (SO) 8(8)--7(7) emission. 
We trace very strong SO emission toward W42-MME using the SMA facility, and 
it is also very strong in the ALMA data having a higher resolution. 
The flattened feature is highlighted in Figure~\ref{fig10}d, and the CO outflow lobes are displayed in Figure~\ref{fig10}e.
In the direction of the proposed infrared jet-like feature, we also detect the noticeable H$^{13}$CO$^{+}$, CH$_{3}$CCH ($K$ =3 transition), and HCO$^{+}$ emission, which is referred to as a small-scale feature. 
The small-scale feature is not located in the direction of the ALMA CO outflow lobes, suggesting that it is unlikely to be a jet. 
In the maps of the continuum, CH$_{3}$OH, SO, and SiO emission, no peak is seen toward the small-scale feature (see also Figures~\ref{fig9}a,~\ref{fig9}d, and~\ref{fig9}e). 
The implication of the observed flattened feature and small-scale structure for the formation of the O-type star is discussed in 
Section~\ref{sec:disc}.
\section{Discussion}
\label{sec:disc}
The present paper deals with a young O-type protostar W42-MME (mass: 19$\pm$4 M$_{\odot}$; luminosity $\sim$4.5 $\times$ 10$^{4}$ L$_{\odot}$). W42-MME is saturated in the {\it Spitzer} 8.0 and 24.0 $\mu$m images, and appears as a point-like source in 
the SOFIA images at 25.2 and 37.1 $\mu$m (see Section~\ref{sec:conta}). 
\citet{dewangan15b} reported a parsec-scale H$_{2}$ outflow driven by this object, and 
employed the high resolution NIR data (resolution $\sim$0\rlap.{$''$}1--0\rlap.{$''$}2) to study its inner circumstellar environment. 
These data sets allowed them to investigate an infrared envelope/outflow cavity (extent $\sim$10640~AU), which surrounds the O-type star and an ionized jet-like feature. 
In this paper, we aim to understand the physical process of mass accumulation in the formation of this young O-type star.
Hence, the findings of \citet{dewangan15b} are used as a basis for further exploring the complex circumstellar environment of the MYSO W42-MME using high resolution ($\sim$0\rlap.{$''$}3--3\rlap.{$''$}5) continuum and spectral line data observed in the sub-mm, mm, and cm regimes.

The ionized clump I4 is found to be close to the position of W42-MME (see Figure~\ref{fig2l}c), but 
new VLA 7 and 13 mm continuum maps do not detect any radio counterpart of the MYSO W42-MME (see Section~\ref{sec:conta}).  
Based on the NIR polarimetric data, the H$_{2}$ outflow axis is parallel to the magnetic
field at the position angle of $\sim$15$\degr$ (see Figure~\ref{fig1}a).

The hot molecular core hosting W42-MME (i.e., MM1) is investigated using the SMA and ALMA molecular line data (see Section~\ref{sec:line}), and is traced with the gas temperature 
of $\sim$38--221~K (see Section~\ref{frottemp1a}). The SMA and ALMA molecular line data also confirm that W42-MME drives a molecular outflow. The ALMA 1.35 mm continuum map shows the presence of an elongated and thermally supercritical filament-like feature (extent $\sim$0.15 pc) containing at least three continuum cores (mass range $\sim$1--4.4 M$_{\odot}$) including MM1 (see Section~\ref{sec:conta}). The elongated filament-feature is also seen in the ALMA continuum map at 865 $\mu$m. As seen in Figures~\ref{fig3}a and~\ref{fig3}b, the ALMA 865 $\mu$m continuum map resolves MM1 into at least two continuum sources MM1a and MM1b.
In the direction of MM1a, at least five continuum sources/peaks (``A--E") are traced within the dusty envelope (extent $\sim$9000~AU), where shocks are investigated in the SiO(8--7) emission. 
The continuum source ``A" associated with W42-MME is found almost at the center of the dusty envelope, and is surrounded by other continuum peaks (``B--E"). A disk-like feature and a small-scale feature are identified toward ``A" using multi-wavelength data (see Section~\ref{sub:inner1}). A variation in the gas temperature (i.e., 38--85~K) toward these features is found in the kinetic temperature map derived using the ALMA CH$_{3}$CCH lines. 

Collectively, these observed features must be interpreted to understand their role in the formation of the O-type star.
\subsection{Signature of an episodic accretion process}
In general, the observed outflows around protostars suggest a disk-mediated accretion process \citep{arce07}. From Figure~\ref{fig7}c, one can identify different H$_{2}$ knots in the northern direction of the H$_{2}$ outflow, showing the shock activity. 
We also find the presence of the SiO(8--7) emission toward the knot (see arrows in Figure~\ref{fig7}c), 
indicating that the shocked gas is associated with the energetic outflow. 
The bipolar structures/lobes centered at W42-MME are traced using the CO(3--2) and SiO(8--7) emission (see Figures~\ref{fig6}b and~\ref{fig6}d). The most prominent H$_{2}$ knot is seen toward the continuum source MM3, which does not appear to be part of the elongated filament-like feature. MM3 has a bow-like appearance in the maps of the HCO$^{+}$(4--3) 
and H$^{13}$CO$^{+}$(4--3) emission, where the diffuse NACO L$'$ emission is also traced. 
The HCO$^{+}$(4--3) and H$^{13}$CO$^{+}$(4--3) emissions show the distribution of the quiescent gas around W42-MME. 
The tip of the bow-like appearance of MM3 is associated with 
the SiO(8--7) emission (see Section~\ref{sec:linez}). 
A water maser is also detected toward MM3, probably showing a signature of shock. 
Additionally, we do not find any extent of the ALMA CO(3--2) or SiO(8--7) outflow lobes toward MM3. 

All these findings suggest the impact of the outflow/jet to the ambient gas around W42-MME, which is an IRc of the 6.7 GHz MME. 
Recently, flaring methanol masers have been found to trace episodic accretion events in young protostars 
\citep[e.g.,][]{bertout89,hirota18,hunter18,macleod18,chen20,liu20,zinchenko20,stecklum21}. 
Hence, our findings appear to show the episodic ejection from W42-MME. 
Such episodic ejection is presumably driven by accretion events, which are known to occur in such 
objects \citep[see][and references therein]{hirota18,zinchenko20,liu20}. 
Together, W42-MME would be a good candidate for monitoring of the methanol (and water) masers in case of a flare. 
\subsection{Disk-like and small-scale structures around W42-MME}
\label{sub:inner2}
Some theoretical simulations \citep[e.g.,][]{mckee03,krumholz09,hosokawa10} predict that massive stars can 
form through disk mediated accretion upto 140 M$_{\odot}$ with very high accretion rates \citep{kuiper10,kuiper13}. 
In addition to the disk-like structure and the outflow cavity, some of the simulations also develop distinct small scale features within a physical scale of about 5000 AU \citep{smith05,krumholz09,peters10a,peters10b,hennebelle11,hennebelle14}. 
These features can be formed by the gravitational instability, 
flashlight effect, jet activity, or radiatively driven Rayleigh-Taylor instability around MYSOs. 

Concerning the validity of the theoretical predictions, direct observational works of the innermost regions of MYSOs are limited. 
Some examples of O-type stars having Keplerian-like disks and outflow have been reported in the literature \citep[see Table~1 in][]{rosen20}, which are AFGL~4176 \citep{johnston15}, G11.92-0.61MM1 \citep{ilee16}, G17.64+0.16 \citep{maud18,maud19}, and IRAS 16547-4247 \citep{Zapata19}. These limited cases favour that MYSOs accrete materials via disk-outflow interaction like their low-mass counterparts \citep[e.g.][]{cesaroni07,zinnecker07,beuther09,beuther13,beltran16}.

The MYSO W42-MME, associated with the continuum source ``A", drives the molecular outflow traced in the H$_{2}$, CO, and SiO emission. 
The self-absorption feature in the HCO$^{+}$ line profile shows infall toward "A". 
From Figure~\ref{fig10}a, the flattened feature and the small-scale structure are evident in the direction of ``A". 
Both these structures are seen within the dusty envelope/outflow cavity. 
The ALMA CO outflow lobes are nearly perpendicular to the flattened feature seen in the H$^{13}$CO$^{+}$ emission (see Figure~\ref{fig10}b). 
Within a scale of 2000~AU, the point-like source traced in the NACO L$'$ image is seen at the center of the flattened feature, suggesting that it could be an accretion disk around the MYSO. 
Figures~\ref{fig10}f,~\ref{fig10}g,~\ref{fig10}h, and~\ref{fig10}i show the moment-1 maps of the SO, H$^{13}$CO$^{+}$, CH$_{3}$OH, 
and NS emission, respectively. The location of the flattened feature is also indicated in each moment-1 map. 
All these moment-1 maps are clipped at the higher value of the cutoff level. 
In Figures~\ref{fig10}f--~\ref{fig10}i, we find a noticeable velocity gradient toward the H$^{13}$CO$^{+}$ flattened feature and 
perpendicular to the outflow. Previously, using the high resolution data (resolution $\sim$0\rlap.{$''$}5) of the dense gas tracers (e.g., CH$_{3}$OH and CH$_{3}$CN), the velocity gradient across the molecular core was suggested as a signature of Keplerian rotation within a rotationally-supported disk \citep[e.g.,][]{zinchenko15}.
 
\begin{figure*}
\epsscale{1}
\plotone{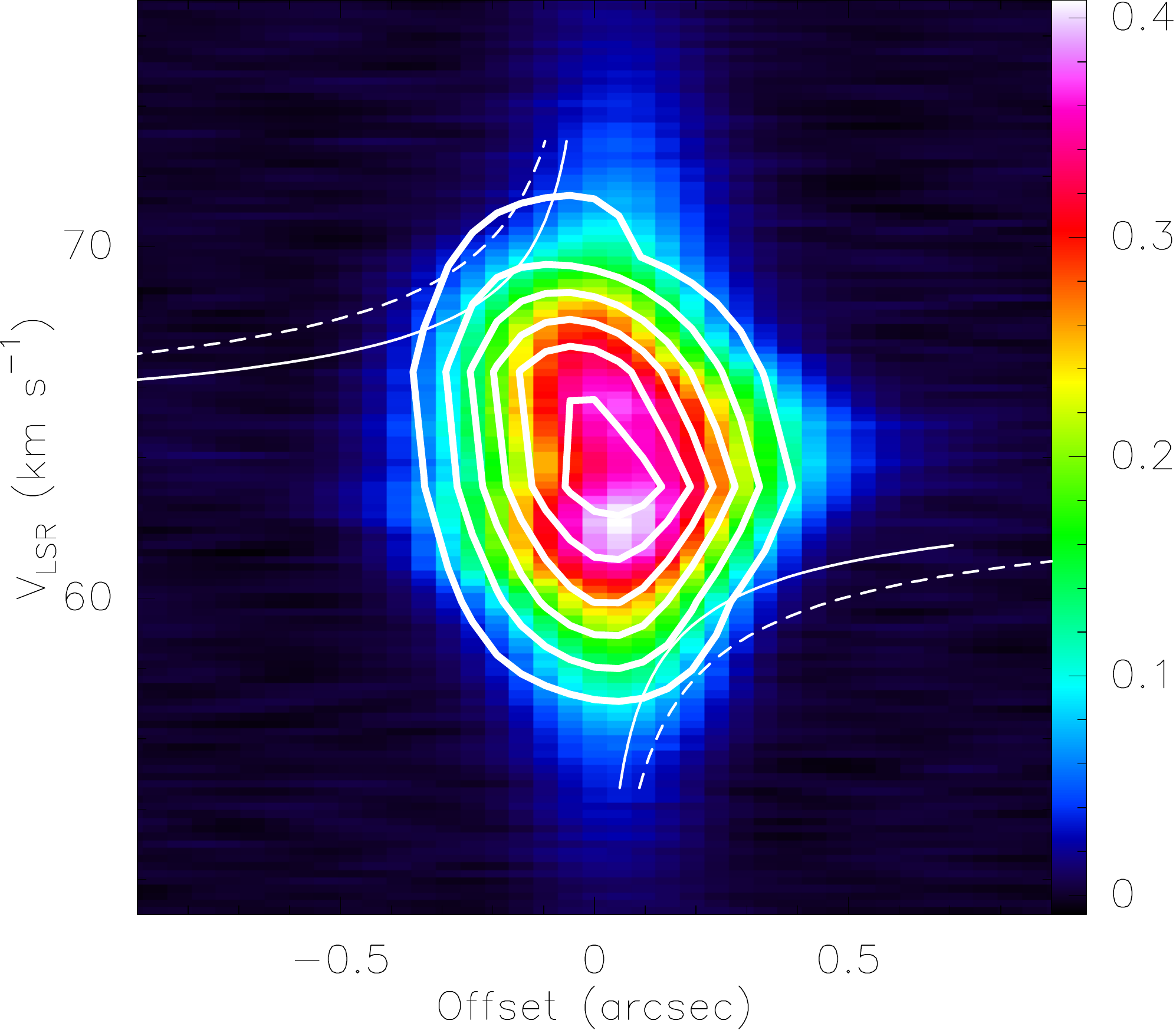}
\caption{Position-velocity diagram along the probable disk in the SO line at the position angle of 132$\degr$ 
across the continuum peak ``A". 
The thick contours (in white) show the CH$_{3}$OH emission.
The curves correspond to Keplerian rotation around the central mass of 
$M~\sin^{2}i$ = 12 M$_{\odot}$ (dashed) and $M~\sin^{2}i$ = 9 M$_{\odot}$ (solid). The color bar displays the SO intensity in Jy beam$^{-1}$.} 
\label{fig11}
\end{figure*}

Figure~\ref{fig11} displays the position-velocity diagram along the probable disk in the SO line 
at the position angle of 132$\degr$ across the continuum peak ``A". 
The contours of the CH$_{3}$OH emission are also shown in Figure~\ref{fig11}. 
The ALMA SO and CH$_{3}$OH data can help us to better characterize the disk kinematics. 
The position-velocity diagram enables us to examine the kinematics of the source ``A", 
and hints the presence of a Keplerian-like rotation of the core ``A". 
However, the resolution of these data is not sufficient for a reliable conclusion \citep[see also][]{zinchenko20,liu20}. 
Using the rotation velocity information (i.e., $M~sin^{2}i$), we compute the dynamical central mass of 
the core to be $\sim$9--12 M$_{\odot}$. Here, the disk inclination is excluded in the calculation. 
The dynamical central mass of the core of 12 M$_{\odot}$ does not properly fit the data (taking into account the 
angular resolution; see Figure~\ref{fig11}), while the value of 9 M$_{\odot}$ can be treated as an upper limit. 
As mentioned earlier, the mass of the continuum source ``A" is estimated to be $\sim$2(1) M$_{\odot}$ at $T_D$ = 40(70) K. 
However, we have temperature estimates up to $\sim$220 K (see Section~\ref{frottemp}). 
Hence, this estimated value of 1 M$_{\odot}$ may show an upper limit of the core mass. 
Additionally, we examine the shape of the position-velocity diagram, where one can find 
the broad ``waist" feature (see Figure~\ref{fig11}). 
Such features in velocity space suggest infalling motions. 
More discussion on the broad ``waist" feature can be found in \citet[][and references therein]{liu20}.

Based on our analysis, we propose two possibilities: 1. To assume that the disk is seen almost face-on. 
However this contradicts the observed elongation in HCO$^{+}$/H$^{13}$CO$^{+}$. But as mentioned, there is no visible elongation in the continuum. We can assume an asymmetry in the HCO$^{+}$ distribution in the disk, so that we see only a half of the disk in these lines. 
Such an assumption is weaker of course but not fully excluded. Another, more natural assumption can be that the bright HCO$^{+}$ emission is related to the cavity walls and does not directly trace the disk.   
2. To assume that we have here a similar situation to that in S255IR-SMA1, where we see a sub-Keplerian rotation accompanied by infall \citep{liu20}. 

The proposed infrared jet-like feature is referred to as a small-scale feature, and is very well-traced in the H$^{13}$CO$^{+}$, CH$_{3}$CCH ($K$ =3 component), and HCO$^{+}$ emission (see Figures~\ref{fig10}a and~\ref{fig10}c). 
Based on the photometric analysis of the NACO NIR images, it was characterized as an ionized jet-like feature. 
However, it does not follow the orientation of the ALMA CO outflow, indicating that it is unlikely to be a jet.
It is seen toward the continuum source ``A". However, there is no peak of the continuum emission or SiO emission found toward the feature. 
This feature does not coincide with the narrower molecular features depicted in H$^{13}$CO$^{+}$. 
It is located within the outflow cavity. 
Interestingly, the CH$_{3}$CCH ($K$ =3 component) peak emission is evident toward the small-scale feature showing gas temperature of $\sim$60--85~K (see Figures~\ref{fig12}c and~\ref{fig12}d). 
There is a temperature gradient evident toward the source ``A" including the small-scale feature. 
Hence, the small-scale feature may be explained as a result of the molecular material being heated by UV radiation from the O-type star. 
\subsection{Scenario for massive star formation}
\label{sub:inner3}
\subsubsection{Hub-filament system in W42}
\label{sssub:inner3}
Several multi-wavelength large-scale surveys reveal the common presence of hub-filament systems in massive star-forming regions \citep[e.g.,][]{Motte+2018,kumar20}. Hence, such configuration is thought to play a significant role in the formation of massive stars. As mentioned earlier, based on the theoretical proposals, massive stars can form via inflow material from very large scales of 1--10 pc (see CA, GNIC, GHC, and Inertial-inflow models), which can be channeled through the molecular cloud filaments. There are two major differences among these scenarios, which are related to 
the driver of the mass flows (turbulence, cloud-cloud collision, etc.) and the existence of the hub-filament structures within molecular clouds. Furthermore, massive stars can also form from the collapse of massive prestellar cores (TC model). 

\begin{figure*}
\epsscale{1}
\plotone{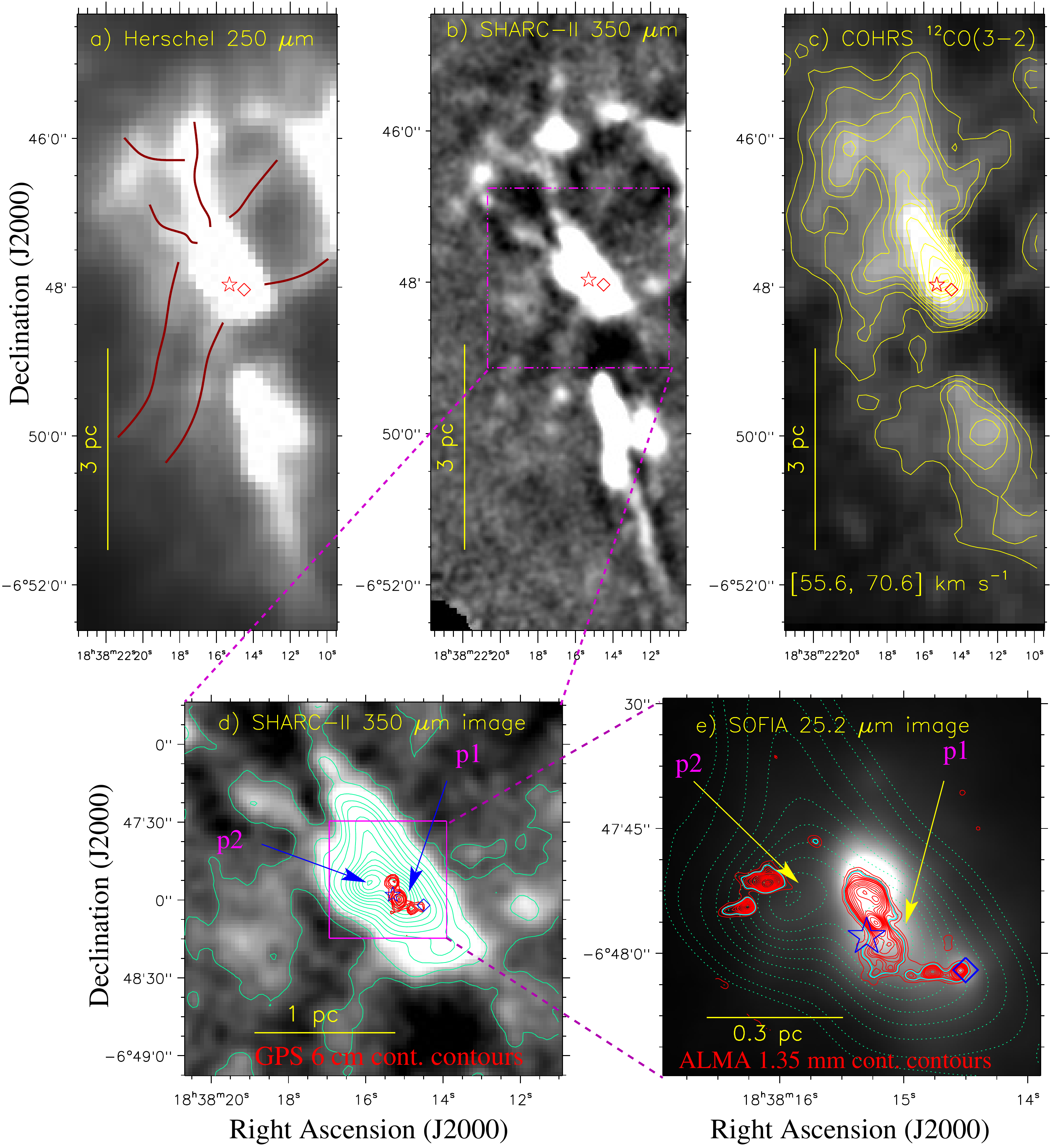}
\caption{Multi-scale picture of W42. A large-scale view of W42 (area $\sim$3\rlap.{$'$}5 $\times$ 8\rlap.{$'$}35; central coordinates: $\alpha_{2000}$ = 18$^{h}$38$^{m}$16\rlap.$^{s}$7, $\delta_{2000}$ = $-$06$\degr$48$'$31\rlap.{$''$}5) using a) the {\it Herschel} image at 250 $\mu$m, b) the SHARC-II 350 $\mu$m continuum image, c) the intensity map and contours of the COHRS $^{12}$CO(3--2) emission integrated over a velocity range of [55.6, 70.6] km s$^{-1}$. The COHRS CO emission contours (in yellow; see Figure~\ref{fig13y}c) are shown with the levels of (0.14, 0.2, 0.25, 0.3, 0.35, 0.4, 0.5, 0.6, 0.7, 0.8, 0.9, 0.98) $\times$ 270.58 K km s$^{-1}$. 
d) The panel shows a zoomed-in view of W42 using the SHARC-II 350 $\mu$m continuum map and contours (see a dotted-dashed box in Figure~\ref{fig13y}b). The SHARC-II 350 $\mu$m continuum contours (in spring green) are displayed with the levels of (0.0032, 0.05, 0.1, 0.15, 0.2, 0.3, 0.4, 0.5, 0.6, 0.7, 0.8, 0.9, 0.98) $\times$ 18.6 Jy beam$^{-1}$. 
The GPS 6 cm continuum contours (in red) are also overlaid on the SHARC-II map (see Figure~\ref{fig1}c). 
e) The panel shows a zoomed-in view of the central part of W42 using the SOFIA 25.2 $\mu$m continuum image (see a solid box in Figure~\ref{fig13y}d). The SOFIA image is also overlaid with the SHARC-II 350 $\mu$m continuum contours (see Figure~\ref{fig13y}d) and the ALMA 1.35 mm continuum contours (in red and cyan; see Figures~\ref{fig2l}a and~\ref{fig2}a)}. In each panel, the positions of a 6.7 GHz MME (diamond symbol) and an O5--O6 star (star symbol) are marked. 
\label{fig13y}
\end{figure*}

\citet{dewangan15c} carefully examined the {\it Herschel} sub-mm images of W42, and identified a hub-filament system in W42. 
Using the {\it Herschel} 250 $\mu$m image, they found parsec-scale filaments, which were radially directed to the denser clump hosting the O5-O6
star and W42-MME. In Figure~\ref{fig13y}a, we present the {\it Herschel} 250 $\mu$m image and highlight several filaments. 
Figure~\ref{fig13y}b displays the SHARC-II 350 $\mu$m image, also revealing the presence of the hub-filament system in W42. 
In Figure~\ref{fig13y}c, we show the intensity map and contours of the COHRS $^{12}$CO(3--2) emission integrated over a velocity range of [55.6, 70.6] km s$^{-1}$, displaying the central molecular condensation containing the O5-O6 star and W42-MME. The COHRS intensity map also confirms the existence of the hub-filament system in W42. Using the SHARC-II 350 $\mu$m image, a zoomed-in view of the hub-filament system in W42 is presented in Figure~\ref{fig13y}d. 
At least two continuum peaks (i.e., p1 and p2) are evident toward the central hub in the SHARC-II 350 $\mu$m image. 
The spatial distribution of the ionized emission traced in the GPS 6 cm continuum map is presented 
in Figure~\ref{fig13y}d, tracing the four ionized clumps (i.e., I1--I4) toward the SHARC-II continuum peak ``p1". 
Figure~\ref{fig13y}e shows a zoomed-in view of the central hub using the SOFIA 25.2 $\mu$m image 
overlaid with the contours of the ALMA band-6 continuum emission at 1.35 mm and the SHARC-II 350 $\mu$m continuum emission. 
The ALMA 1.35 mm continuum emission is the same as presented in Figure~\ref{fig2l}a.  
In Section~\ref{sec:conta1x}, we already discussed the GPS~6~cm continuum map and the SOFIA mid-infrared image. 
The locations of the O5-O6 star and W42-MME are spatially seen toward the SHARC-II continuum peak ``p1", 
where the mid-infrared emission is prominently evident. 
Using the VLT/NACO adaptive-optics K$_{s}$-band and L$^{\prime}$-band images, \citet{dewangan15c} 
examined the inner environment of the O5-O6 star (see Figure~2 in their paper). 
However, in the direction of the SHARC-II continuum peak ``p2", at least two 1.35 mm continuum peaks are found, 
and are not associated with any mid-infrared and radio emission. 
No K-band sources are also seen toward these two peaks (not shown here). 
Based on these results, the two 1.35 mm continuum peaks located toward 
the SHARC-II continuum peak ``p2" may be candidates of massive prestellar cores, which 
deserve further investigation with the molecular line data. 
However, a detailed study of massive prestellar cores is beyond the scope of this present work. 

Overall, on a large-scale picture of W42 (i.e., 3 pc $\times$ 3 pc), several mm continuum cores, hosting massive stars, are investigated inside a central hub, which is surrounded by several parsec scale filaments. It implies that the material to form massive stars (including O5-O6 star and W42-MME) in W42 appears to be collected through the filaments. In a given massive star-forming region, the presence of a hub-filament configuration may hint the applicability of the GNIC scenario \citep{Motte+2018}, which includes the flavours of the CA and GHC models.
Hence, the GNIC scenario seems to be applicable in W42. 
\subsubsection{Formation process of the massive O-type star W42-MME}
\label{sssub:inner3x}
The massive O-type star W42-MME is embedded in the dust continuum clump, which appears to be grown by gaining the inflowing material that is channeled by the filaments (see Section~\ref{sssub:inner3} for more discussion). 

In the direction of W42-MME, several continuum sources over a scale of 0.27 pc are evident in the ALMA continuum map at 865 $\mu$m (resolution $\sim$0\rlap.{$''$}3); six of these are MM1a, MM1b, MM2, MM3, MM4, and MM5. 
Over a scale of 0.1 pc, three sources, MM1a, MM1b, and MM2, are seen inside a common contour level of the ALMA continuum emission at 865 $\mu$m. In the direction of the continuum source MM1a, a dusty envelope (extent $\sim$9000~AU) containing 
at least five continuum sources/peaks (``A--E") is seen in the ALMA continuum map at 865 $\mu$m. 
The continuum source ``A" associated with W42-MME is found almost at the center of the dusty envelope, and is surrounded by other continuum peaks (``B--E"). Hence, the mass reservoir available for the birth of a single O-type star associated with the continuum 
source ``A" seems plentiful. Here one can keep in mind that the continuum source MM1a (mass $\sim$2--3.8 M$_{\odot}$; see Table~\ref{tab2}) 
is not massive enough to form a massive star, but it hosts W42-MME and is associated with a hot molecular core. 
In this relation, we can suggest that cores forming massive stars do not accumulate all 
the mass before core collapse, but instead, cores and embedded protostars gain mass simultaneously 
\citep[e.g.,][]{zhang09,wang11,wang14,sanhueza19,svoboda19}. It is consistent with the GNIC scenario and/or the CA and the GHC scenarios (see Section~\ref{sssub:inner3}).

In the maps of the H$^{13}$CO$^{+}$ and HCO$^{+}$ emission, within a scale of 10000 AU, narrow molecular structures surrounding the continuum source ``A" are evident toward the dusty envelope.
The positions of the continuum sources (i.e., B, C, and D) are spatially found toward these narrow structures of
the H$^{13}$CO$^{+}$ emission. SiO outflow lobes are spatially concentrated toward ``A", while shocks are also seen toward B and C in the SiO(8--7) emission (see Figure~\ref{fig7}c). 
Mass estimation of these sources will not be accurate because they are influenced by shocks. 
The dusty envelope or outflow cavity (extent $\sim$9000~AU) is associated with shocks as traced in the SiO(8--7) emission.
Dynamical mass of the core ``A" is estimated to be $\sim$9 M$_{\odot}$. 
The analysis of the H$^{13}$CO$^{+}$ profile shows the domination of the non-thermal pressure and supersonic non-thermal motions around the continuum source ``A" (see Section~\ref{llsub:inner1}).

In recent years, high resolution observations ($\sim$1000s AU scale) of accreting MYSOs indicate that massive stars can form through infall 
from a surrounding envelope. Furthermore, the growth of an accretion disk facilitates an accretion flow onto the central object \citep[see latest review article by][for more details]{rosen20}. 

Our observational outcomes also favour the onset of the disk-mediated accretion process in the MYSO W42-MME. 
We also propose that the core ``A" accretes material from the envelope as well as from the immediate surrounding cores.
\section{Summary and Conclusions}
\label{sec:conc}
We observed in the sub-mm, mm, and cm regimes the dust, ionized emission, and molecular gas
surrounding the MYSO W42-MME (mass = 19$\pm$4 M$_{\odot}$; luminosity $\sim$4.5 $\times$ 10$^{4}$ L$_{\odot}$) using the ALMA, SMA, and VLA interferometric facilities (resolution $\sim$0\rlap.{$''$}3--3\rlap.{$''$}5). 
Our conclusions are as follows.\\
$\bullet$ An elongated filament-like feature (extent $\sim$0.15 pc) is investigated in the ALMA 1.35 mm continuum map, and is characterized as a thermally supercritical 
filament. Three continuum cores (mass range $\sim$1--4.4 M$_{\odot}$) are seen toward this feature, 
and one of these cores (i.e., MM1; mass$\sim$4.4 M$_{\odot}$) hosts the MYSO W42-MME.\\ 
$\bullet$ The ALMA 865 $\mu$m continuum map reveals at least five continuum sources/peaks (``A--E") within a dusty envelope (extent $\sim$9000~AU) toward MM1, where shocks are traced in the SiO(8--7) emission.
The continuum source ``A" associated with W42-MME is found almost at the center of the dusty envelope, and is surrounded by other continuum peaks (``B--E"). \\
$\bullet$ The kinetic temperature map derived using the ALMA CH$_{3}$CCH lines shows the presence of a temperature gradient toward the continuum source ``A". The gas temperature ranges from 38 to 85~K. \\ 
$\bullet$ The SMA and ALMA facilities have detected dense/hot gas tracers ($^{13}$CS(5--4), HC$_{3}$N(24--23), CH$_{3}$CCH, CH$_{3}$CN, CH$_{3}$OH(5$_{1,4}$--4$_{2,2}$), and CH$_{3}$OH(4$_{1,3}$--3$_{0,3}$)) and shock tracer SiO toward W42-MME. 
Based on the rotational diagram analysis of several transitions of the ALMA band-6 CH$_{3}$CN emission, 
the rotational temperature is estimated to be $\sim$220~K. Our analysis confirms the presence of a hot molecular core associated 
with W42-MME.\\ 
$\bullet$ A molecular outflow is traced in the SMA CO(2--1) line data, and is centered at the continuum peak MM1.
The ALMA CO(3--2) line observations resolve the bipolar northeast-southwest outflow associated with the 
continuum source ``A", which is distributed within a scale of 10000 AU. The bipolar outflow is also traced in the ALMA SiO(8--7), which is spatially concentrated toward 
the continuum source ``A". \\
$\bullet$ The molecular multiline data trace the dense cavity walls toward the dusty envelope 
around W42-MME at below 10000 AU, where shocks are traced in the SiO(8--7) emission.\\
$\bullet$ The continuum source MM3 has a bow-like appearance, and is associated with the H$_{2}$, H$_{2}$O maser, and SiO(8--7) emission.
Very strong intensities of the HCO$^{+}$(4--3) and H$^{13}$CO$^{+}$(4--3) emission are observed toward MM3, which is located in the northern side of the dusty envelope. It seems that MM3 possibly originated from a previous ejection event from the MYSO W42-MME.
In other words, there is a signature of episodic ejections from W42-MME, favouring a disk-mediated variable accretion event.\\
$\bullet$ Based on the velocity gradient seen in the ALMA multiline data (e.g., HCO$^{+}$(4--3), SO, CH$_{3}$OH, and NS emission), the dynamical central mass of the core hosting W42-MME is computed to be $\sim$9 M$_{\odot}$. No disk inclination is considered in the calculation.\\
$\bullet$ Within a scale of 2000 AU, the flattened/elongated feature is investigated in the continuum source ``A" using the H$^{13}$CO$^{+}$(4--3) emission, and is perpendicular to the orientation of the ALMA CO outflow. A noticeable velocity gradient is also observed across the flattened/elongated feature in the CH$_{3}$OH, H$^{13}$CO$^{+}$, SO, and NS maps.\\
$\bullet$ In the direction of the continuum source ``A", the position-velocity maps of the SO, CH$_{3}$OH, HCO$^{+}$(4--3), and NS emission 
hint the existence of a Keplerian-like rotation within a rotationally-supported disk (mass $\sim$1 M$_{\odot}$). The resolution of the data is not enough for a firm conclusion.\\
$\bullet$ An asymmetric self-absorbed line profile of an optically thick HCO$^{+}$ line supports the signatures of infall toward the continuum source ``A". The position-velocity map of the SO emission reveals the presence of a wider ``waist" like feature, which shows the signature of infalling motions in the source ``A".\\
Overall, our observational findings show the disk-mediated accretion process in the MYSO W42-MME. 
We also suggest that the core hosting W42-MME appears to gain mass from the envelope and also from the immediate surrounding cores.
\acknowledgments  
We thank the anonymous reviewer for several useful comments and 
suggestions, which greatly improved the scientific contents of the paper.  
The research work at Physical Research Laboratory is funded by the Department of Space, Government of India. 
I.I.Z., P.M.Z. and A.G.P. acknowledge the support by the Russian Science Foundation (grant No. 17-12-01256). SY Liu acknowledges the support from Ministry of Science and Technology through the grant MOST 109-2112-M-001-026. 
DKO acknowledges the support of the Department of Atomic Energy, Government of India, under project Identification No. RTI 4002.
This paper makes use of the following ALMA data: ADS/JAO.ALMA\#2018.1.01318.S and ALMA archive data: ADS/JAO.ALMA\#2019.1.00195.L. 
ALMA is a partnership of ESO (representing its member states), NSF (USA) and NINS (Japan), together with NRC (Canada), MOST and ASIAA (Taiwan), and KASI (Republic of Korea), in cooperation with the Republic of Chile. The Joint ALMA Observatory is operated by ESO, AUI/NRAO and NAOJ. In addition, publications from NA authors must include the standard NRAO acknowledgement: The National Radio Astronomy Observatory is a facility of the National Science Foundation operated under cooperative agreement by Associated Universities, Inc.
This work is based [in part] on observations made with the {\it Spitzer} Space Telescope, which is operated by the Jet Propulsion Laboratory, California Institute of Technology under a contract with NASA. 
This publication makes use of data from FUGIN, FOREST Unbiased Galactic plane Imaging survey with the Nobeyama 45-m telescope, a legacy project in the Nobeyama 45-m radio telescope. 
\begin{table*}
\setlength{\tabcolsep}{0.1in}
\centering
\caption{Table lists names, positions, flux densities, deconvolved FWHM$_{x}$ \& FWHM$_{y}$, and masses of 
the continuum sources traced at 865 $\mu$m (see Figure~\ref{fig3}a). 
The uncertainty in the mass estimate can be $\sim$20\% and at largest $\sim$50\%.}
\label{tab2}
\begin{tabular}{lccccccccccccc}
\hline 										    	        			      
  Name           &  $\alpha_{2000}$ & $\delta_{2000}$         & Total Flux &FWHM$_{x}$ $\times$ FWHM$_{y}$ & Mass (M$_{\odot}$) & Mass (M$_{\odot}$)  \\ 
                 &  (h m s)         &  ($\degr$ $'$ $''$)     &     (mJy)  & ($''$ $\times$ $''$)    &  at $T_D$ = 40 K      &  at $T_D$ = 70 K   \\ 
\hline 
   MM1a  &  18:38:14.54  &$-$06:48:02.0  & 120.1 &   0.63 $\times$ 0.99    &	3.8   &   2.0  \\
   MM1b  &  18:38:14.64  &$-$06:48:02.9  & 14.9  &   0.69 $\times$ 0.61    &	0.5   &   0.2  \\
   MM2   &  18:38:14.76  &$-$06:48:02.1  & 80.7  &   0.86 $\times$ 0.67    &	2.5   &   1.3  \\
   MM3   &  18:38:14.67  &$-$06:47:57.8  & 42.9  &   0.68 $\times$ 1.52    &	1.3   &   0.7  \\
   MM4   &  18:38:14.96  &$-$06:48:02.6  & 14.6  &   0.83 $\times$ 0.39    &	0.5   &   0.2  \\
   MM5   &  18:38:14.16  &$-$06:48:03.5  & 29.2  &   1.21 $\times$ 1.40    &	0.9   &   0.5  \\
\hline          		
\end{tabular}			
\end{table*}			
\clearpage

\end{document}